\documentclass[]{spie}  
\usepackage[]{graphicx}
\usepackage{amsmath}

\title{PILOT: a balloon-borne experiment to measure the polarized FIR emission of dust grains in the interstellar medium} 

\author{
R. Misawa\supit{a}  and
J-Ph. Bernard\supit{a} and
P. Ade\supit{e} and
Y. Andr\'e\supit{g} and
P. deBernardis\supit{d} and
M. Bouzit\supit{b} and
M. Charra\supit{b} and
B. Crane\supit{b} and
 J.P. Dubois\supit{b} and
C. Engel\supit{a}  and
M. Griffin\supit{e}  and
P. Hargrave\supit{e}  and
B. Leriche\supit{b} and
Y. Longval\supit{b} and
S. Maes\supit{a} and
C. Marty\supit{a} and
W. Marty\supit{a} and
S. Masi\supit{d} and
B. Mot\supit{a} and
 J. Narbonne\supit{a}  and
F. Pajot\supit{b} and
G. Pisano\supit{e} and
N. Ponthieu\supit{f} and
I. Ristorcelli\supit{a} and
L. Rodriguez\supit{c} and
G. Roudil\supit{a} and
M. Salatino\supit{d} and
G. Savini\supit{e} and
C. Tucker\supit{e}
\skiplinehalf
\supit{a}Institut de Recherche en Astrophysique et Planetologie (IRAP),
9 Av du Colonel Roche, BP 4346, 31028 Toulouse cedex 4; \\
\supit{b} Institut d'Astrophysique Spatiale (IAS), B\^at 121, Universit\'e
Paris XI, Orsay, France; \\
\supit{c}CEA/Saclay, 91191 Gif-sur-Yvette Cedex, France; \\
\supit{d}Universita degli studi di Roma "La Spienza", Dipartimento di Fisica, 
P.le A. Moro, 2, 00185, Roma, Italia; \\
\supit{e}Department of Physics and Astrophysics, PO BOX 913, Cardiff University, 
5 the Parade, Cardiff, UK; \\
\supit{f}Grenoble University, Grenoble, France; \\
\supit{g}Centre National des Etudes Spatiales, DCT/BL/NB, 18 Av. E.
Belin, 31401  Toulouse, France; \\
}

\authorinfo{Further author information: (Send correspondence to J.P.B.)\\J.P.B.: E-mail: Jean-Philippe.Bernard@irap.omp.eu, Telephone: 33 (0)5 61 55 75 38\\  R.M.: E-mail: rmisawa@irap.omp.eu, Telephone: 33 (0)5 61 55 75 38}

\begin{document} 

\newcommand{\Archeops}{\mbox{\sc{Archeops }}}
\newcommand{\Boomerang}{\mbox{\sc{Boomerang }}}
\newcommand{\Cobe}{\mbox{\sc{Cobe }}}
\newcommand{\Dirbe}{\mbox{\sc{Dirbe }}}
\newcommand{\Firas}{\mbox{\sc{Firas }}}
\newcommand{\Cbi}{\mbox{\sc{Cbi }}}
\newcommand{\Dasi}{\mbox{\sc{Dasi }}}
\newcommand{\Dmr}{\mbox{\sc{Dmr }}}
\newcommand{\Hfi}{\mbox{\sc{Hfi }}}
\newcommand{\Iras}{\mbox{\sc{Iras }}}
\newcommand{\Wmap}{\mbox{\sc{Wmap }}}
\newcommand{\Maxima}{\mbox{\sc{Maxima }}}
\newcommand{\Planck}{\mbox{\sc{Planck }}}
\newcommand{\TopHat}{\mbox{\sc{TopHat }}}
\newcommand{\Vsa}{\mbox{\sc{Vsa }}}
\newcommand{\Pilot}{\mbox{\sc{PILOT }}}

\newcommand{\mic}{\,{\rm \mu m} }
\newcommand{\degr}{{\rm ^o} }
\newcommand{\etal}{et al.}

\newcommand{\polang}{p}
\newcommand{\polfrac}{\psi}

\newcommand{\archpol}{ArchPol}

\def\ApJ{{\sl Astrophys. J.}}
\def\ApJL{{\sl Astrophys. J. Lett.}}
\def\AA{{\sl Astron. Astrophys.}}
\def\AASS{{\sl Astron. Astrophys. Supp. Ser.}}
\def\NCA{{\sl Nuovo Cimento}}
\def\NIM{{\sl Nucl. Instrum. Methods}}
\def\NIMA{{\sl {Nucl. Instrum. Methods} A}}
\def\NPB{{\sl {Nucl. Phys.} B}}
\def\PLB{{\sl {Phys. Lett.}  B}}
\def\PRL{{\sl Phys. Rev. Lett.}}
\def\PRD{{\sl {Phys. Rev.} D}}
\def\ZPC{{\sl {Z. Phys.} C}}
\def\PR{{\sl Phys. Rept.}}
\def\NewA{{\sl New Astron.}}
\def\AJ{{\sl Astron. J.}}
\def\CASP{{\sl Comm. Ap. Sp. Phys.}}
\def\MNRAS{{\sl Month. Not. Roy. Ast. Soc.}}

\def\Cl{C_\ell}

\include{journals_defs}

\providecommand{\sorthelp}[1]{}
\def\GHz{\ifmmode $\,GHz$\else \,GHz\fi}

\maketitle 

\begin{abstract}

Future cosmology space missions will concentrate on measuring the
polarization of the Cosmic Microwave Background, which potentially
carries invaluable information about the earliest phases of the evolution
of our universe.  Such ambitious projects will ultimately be limited by
the sensitivity of the instrument and by the accuracy at which
polarized foreground emission from our own Galaxy can be subtracted
out.  We present the \Pilot balloon project which will aim at
characterizing one of these foreground sources, the polarization of
the dust continuum emission in the diffuse interstellar medium. The
\Pilot experiment will also constitute a test-bed for using
multiplexed bolometer arrays for polarization measurements.
We present the results of ground tests obtained just before the first flight of the instrument.

\end{abstract}



\section{INTRODUCTION}
\label{sec:intro}  

Current theories predict that, much before recombination, well within
the first second after the Big-Bang, the universe underwent a
period of extremely fast expansion called inflation.  During this
phase, the size of the universe grew from microscopic to
macroscopic almost instantly.  The inflation theory provides a natural
explanation to several paradoxes of the big-bang theory, and is
therefore favoured by many cosmologists and particle physicists.
Observations of the intensity of the CMB anisotropies cannot reveal the
inflation phase, since all information from periods prior to the last
scattering surface was lost.  If they were not so difficult to detect
by nature, density waves generated during the inflation would bring us
invaluable information about these very early phases.  Luckily, the
interaction between the quadrupole anisotropies induced by primordial
gravitational waves and the CMB photons at the last scattering surface
is expected to have polarized a tiny fraction of the CMB light (called
B-modes).  Measuring this
polarized signal would allow cosmologists to look behind the curtain of
the last scattering surface and would alleviate degeneracies in the
present cosmological parameter determination.  However, the CMB B-modes
are extremely weak.  Depending upon the energy scale at which inflation
actually occurred, they may not even be observable in the foreseeable
future.  In any case, their detection promises to be the next major
challenge of observational cosmology and the next major step forward
for cosmology and fundamental physics.

Several mission concepts for future satellite missions to measure
B-modes are already under study in the USA and Europe (such as EPIC,
Einstein Probe for Inflationary Cosmology or CMBpol).  The main
limitations to CMB polarization measurements will be the sensitivity of
the measurements and the confusion with unrelated foreground polarized
emission.  The current observation sensitivity in the microwave domain
is already limited by photon noise from the CMB radiation itself, so
that it can only be improved by massively increasing the number of
detectors.  This will require using multiplexed bolometer filled arrays
of very large format.  Such devices are slowly becoming available.
Note however that systematic effects induced by the instrument itself
will also have to be controlled down to very low levels in order to
achieve ultimate sensitivities.  Confusion arises from polarized
emission produced by several other astronomical sources.  First, the quadrupole
anisotropies caused by the electron motion near density perturbations
at the time of decoupling (so called scalar perturbations) is expected
to produce additional CMB polarization, which is not directly linked to
density waves and inflation.  Unlike for B-modes, the associated
polarized signal (called E and TE-modes) is expected to be correlated
with the density, and therefore with the CMB intensity fluctuations.
Also, they are expected to have a different parity than that of the
B-modes, which could allow separation of the various components,
despite the fact that B-modes are much weaker than the E and TE-modes.
A second source of foreground is the polarized emission originating
within our own Galaxy, either from sub-micron size dust particles
(polarized dust emission) or gas particles (synchrotron radiation)
interacting with the magnetic field of our Galaxy.  The 3-year WMAP
results \cite{page2007} have shown that, already for E and TE
modes, correcting for the contribution of polarized foreground emission
is a crucial step towards an accurate measurement of CMB polarization.
At the accuracy levels required to detect B-modes, no piece of sky will
be clean, even at high galactic latitude, and polarized foregrounds
will have to be precisely understood and removed before any
cosmological interpretation of the data can take place.  Although a
very accurate knowledge and understanding of polarized foreground
signals will be needed, their amplitude and characteristics are
presently very poorly constrained by observations.

Many experiments are currently engaged in the measurement of
polarized CMB and the search for B-modes, either from the ground (e.g.
QUaD, BICEP, Clover, BRAIN, QUIET, \ldots) or from balloons (e.g.
Bar-Sport, EBEX, Spider, \ldots). However, none of them is actually designed
specifically to address the foreground contamination issue.
Here, we present the \Pilot experiment, which will tackle the problem
of precisely measuring the polarized emission from dust grains in the
diffuse InterStellar Medium (ISM).

\section{Dust Polarization}

The ISM is populated by very tenuous gas
(essentially neutral Hydrogen, but also more complex molecules) and
dust grains of sub-micron size which are composed of solid, amorphous
silicates and graphite.  These grains absorb starlight mainly in the UV
and visible, are heated to 15-30 K in the diffuse ISM and radiate
continuum emission in the Far-Infrared to Sub-millimeter wavelength
range, where they dominate the total sky emission.  Dust particles are
believed to have somewhat elongated shapes.  In the presence of the
faint magnetic field (a few micro-gauss) that pervades in the ISM,
several processes contribute to make dust grains rotate about their
minor axis, which then partially aligns with respect to the magnetic
field lines.  This partial alignment of dust grains causes preferential
absorption of starlight along the major axis of the grains, so that
unabsorbed star light appears slightly polarized along the field
direction.  Similarly, emission by partially aligned dust grains causes
a fraction of the thermal emission to be linearly polarized in a
direction orthogonal to the magnetic field line, as projected on the
sky.


For a long time, dust emission polarization measurements remained very scarce and
suffered from large biases.  Ground-based observations in the visible
allowed to measure the polarization in absorption toward stars
\cite{fosalba2002}, but were restricted to selected directions on the sky
where bright stars are present.  Due to their low sensitivity,
polarization measurements in emission obtained from large telescopes on
the ground allowed to probe only the brightest regions of the sky, in the
immediate vicinity of star forming regions.  These regions show
features which are specific to extreme environments where stars form,
and which are not representative of the physical effect which will
dominate in more diffuse regions.  In particular, it is possible that the
increased gas density in those regions somewhat disrupts the grain
alignment with the magnetic field, which could explain the saturation of
the polarization fraction observed with rising column density in those
regions and the very low polarization fractions (of the order of $2\%$)
observed there \cite{Hildebrand1999}.

The balloon experiment \Archeops \cite{Benoit2004a} measured the
dust emission polarization at 353 GHz over a large fraction of the
milky way \cite{Benoit2004}.  A significant polarized signal was
detected toward several individual regions of the Galactic plane, where
the intensity is high enough for its polarization to be measured.  They
correspond to individual Molecular Clouds (MCs).  The polarization
degree towards those regions is larger than $5\%$, stronger than
previously expected, and is sometimes larger than any region of the sky
already measured (consistent with polarization degrees up to $10\%$).
Note that these figures are average values taken over fairly large
regions of the sky, of the order of 3-50 square degrees.  Note also
that currently available techniques from the ground do not presently
allow the detection of polarization over such large areas with moderate
to low intensity.
In principle, they may correspond to regions where the geometry of the magnetic field orientation is very
coherent and close to the plane of the sky over large volume areas.
They could also correspond to molecular clouds where the dust
properties vary and grain alignment is favored.  Despite its success,
the \Archeops limited sensitivity did not allow a direct determination
of the polarization towards diffuse regions at higher galactic
latitude.

Recently, the Planck satellite measurements have finally revealed
the structure of dust polarization over large regions of the sky.  A
tight correlations between polarization in VIS extinction towards
selected stars and the corresponding submm emission
(\cite{planck2014-XXI}) confirms that the same grains are
producing both. The large ratio of submm to VIS polarization
challenges current dust models.  Several regions were identified with
polarization fractions at 353 GHz up to almost $20\%$
(\cite{planck2014-XIX}). These highly polarized regions may
correspond to regions where the geometry of the magnetic field
orientation is very coherent and close to the plane of the sky over
large volume areas. Their existence confirms that dust in diffuse
regions can be efficiently aligned and intrinsically polarized to much
higher fractions than previously expected. The Planck data also
revealed that the direction of the apparent dust emission
polarization, which traces the sky projection of the magnetic field
averaged over the LOS, is generally homogeneous in large
polarization fraction regions, and rotates abruptly in low
polarization fraction regions. The high rotation regions draw
spectacular filamentary features. These results indicate that the 3D
geometry of the magnetic field along the LOS is a key parameter
determining the apparent dust polarization fraction. This was
confirmed by a comparison with predictions from MHD simulations that
reproduces the observations (\cite{planck2014-XX}). Inspection of
the spectral dependence of the polarization fraction over the Planck
spectral range (\cite{planck2014-XXI}) indicates a weak but steady
decline of the polarization fraction with frequency that could reflect
either dust properties, such as metallicinclusions in dust grains, or
progressive reduction by unproperly understood additional emission
sources, such as spinning dust or free-free emission.  Finally, the
expected correlation between dust and synchrotron polarized emission,
although present, is rather poor (\cite{planck2014-XIX}),
suggesting that dust and high energy cosmic rays generally sample
quite different regions of the LOS and magnetic field directions.

\section{PILOT DESCRIPTION}
\label{sec:PILOTDESCRIPTION}

\Pilot (Polarized Instrument for the Long-wavelength Observations of
the Tenuous ISM: http://pilot.irap.omp.eu , is a
project of a balloon-borne astronomy experiment to study the
polarization of dust emission in the diffuse ISM
in our Galaxy.

The \Pilot instrument will allow observations in two photometric
channels at wavelengths $240\mic$ (1.2 THz) and $550\mic$ (545 GHz) at
an angular resolution of a few arcminutes.  We will make use of large
format bolometer arrays, newly developed for the PACS instrument on
board the Herschel satellite.  With 1024 detectors per photometric
channels and photometric bands optimized for the measurement of dust
emission, \Pilot is likely to become the most sensitive experiment for
this type of measurements.  The proposed method to measure
polarization using filled arrays has been validated
by end-to-end simulations.  The \Pilot experiment will take advantage
of the large gain in sensitivity allowed by the use of large format,
filled bolometer arrays at frequencies more favorable to the detection
of dust emission.  Its sensitivity will surpass that of \Archeops and
even that of Planck (by a factor about 20-30) for a given column
density of interstellar matter.  It will prefigure that of future
cosmology experiments. Note that \Pilot will
be the only instrument able to measure the polarization properties of
dust emission in the FIR, since none of the Herschel satellite
instrument will be sensitive to polarization.

\begin{figure}[htb]
\centering
\includegraphics[width=15cm]{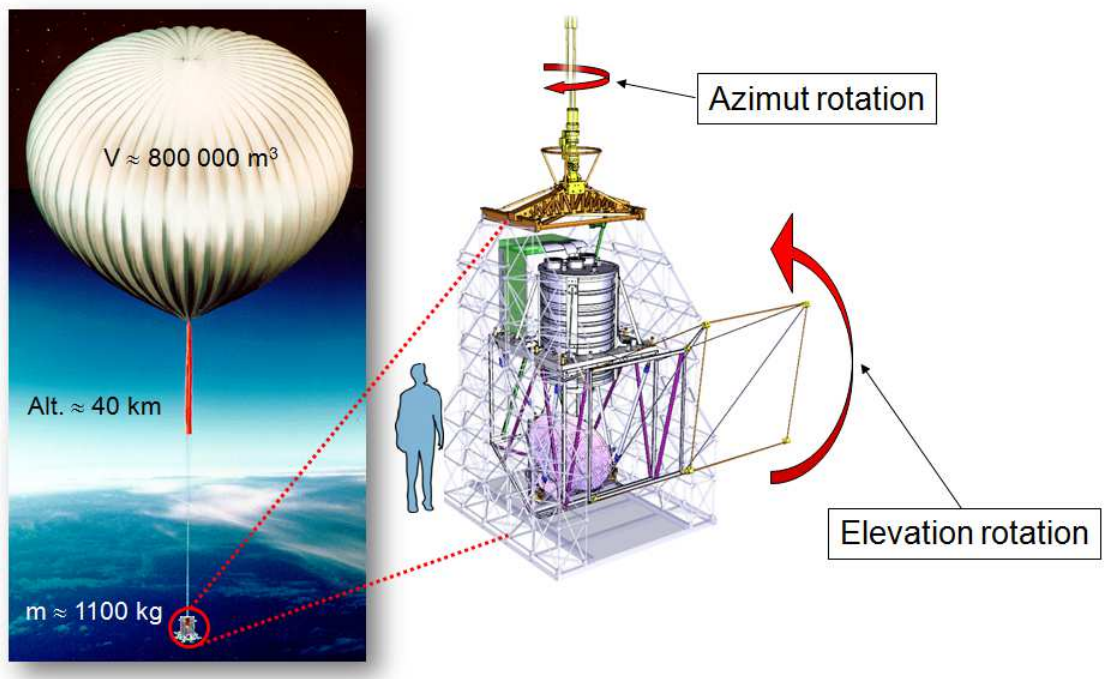}
\caption{\footnotesize Schematic view of the \Pilot gondola. The cryostat
(cylinder on top) and the primary mirror are
attached to the pointed load, which can rotate around its elevation
axis in order to change the elevation of the submm optical axis.
Motion around the flight chain will be ensured by the azimuth swivel. Optical baffles and thermal protection screens
surrounding the experiment are not shown.
\label{fig:gondola}}
\end{figure}


\subsection{Science Goal}


The observations of \Pilot have two major scientific objectives.  On
the one hand, they will allow us to constrain the large scale geometry
of the magnetic field in our Galaxy and to study in details the
alignment properties of dust grains with respect to the magnetic field.
In this domain, the measurements of \Pilot will be complementary with
those of Planck at longer wavelengths.  In particular, they will bring
information at a higher angular resolution, which is critical in
crowded regions such as the Galactic plane.  They will allow us to
better understand how the magnetic field is shaping the ISM material on
large scale in molecular clouds, and the role it plays in the
gravitational collapse leading to star formation.  On the other hand,
the \Pilot observations will allow us to measure for the first time the
polarized dust emission towards the most diffuse regions of the sky,
where the measurements are the most easily interpreted in terms of the
physics of the dust.  In this particular domain, \Pilot will play a
role for future CMB missions similar to that played by the Archeops
experiment for Planck.  The results of \Pilot will allow us to gain
knowledge about the magnetic properties of dust grains and to the
structure of the magnetic field in the diffuse ISM that will be
necessary to a precise foreground subtraction in future CMB
measurements.  For instance, one of the major objectives is to test if
the dust polarization degree changes with wavelength, as some ground
observations towards dense regions seem to indicate
(see \cite{Hildebrand2009}).  Dust models which invoke the existence of
two separate dust components at different temperatures in the ISM to
explain the sub-mm emission of our Galaxy
(e.g. \cite{finkbeiner1999}) are likely to produce such changes, since
different types of grains dominate the FIR and submm emission.
However, most recent models advocating for the submm emission being
produced by low energy transitions in the amorphous material composing
the grains (\cite{Meny2007}) are expected to produce very little
variations of the polarization properties with wavelength, since the
same dust grains dominate the emission over the whole spectrum.  The
\Pilot measurements, combined with those of Planck at longer
wavelengths, will therefore allow us to further constrain the dust
models.  The outcome of such studies will likely impact the
instrumental and technical choices for the future space missions
dedicated to CMB polarization.

\Pilot will not only detect polarized dust emission, but will also
measure its total intensity with high accuracy.  When combined with the
IRAS and AKARI all-sky surveys at $100\mic$ and $160\mic$ respectively,
the large scale survey performed with the \Pilot experiment at
$240\mic$
will be important in deriving the dust temperature, which is a
critical quantity for all dust studies. In combination with the \Planck
data at longer wavelengths \cite{planck2013-p01,planck2013-XVII,planck2013-p06b}, and the \Pilot data at $500\mic$,
this will be used to study the dust emissivity slope over large areas of
the sky.
Note that accurate determinations of the dust temperature distribution
are obviously be possible with the Herschel satellite data, but
have to be focused on limited sky areas, since the largest surveys
with Herschel cover a small fraction of
the sky, with the Herschel galactic plane survey (HiGal) covering only 2
degrees in galactic latitudes \cite{Molinari2010}.  Scientific by-products of the
\Pilot experiment will also include a large point-source catalog
(several hundred sources should be detected per hour).

\subsection{Instrument description}

\Pilot is a balloon-borne experiment being designed to fly at a ceiling
altitude of around 40 km (4 hPa pressure) in the stratosphere.  The
experiment will be carried by a generic CNES stabilized gondola with
altazimuthal coarse pointing control, as shown in
Fig.\,\ref{fig:gondola}.
Mapping of the sky will be accomplished by rotating the gondola over a
large azimuth range ($\rm \pm 30^{o}$) at constant elevation, in order to
reduce the residual atmospheric contribution.
The elevation of the pointed payload can range from $20^{o}$ to
$60^{o}$.  The fine attitude of the instrument and the effective
pointing directions will be constructed a posteriori, using the signal
from a fast, large format CCD stellar sensor co-aligned with the sub-mm
axis.  The fastest ($1.2^{o}$/s) rotation speed for the azimuth
scanning is a compromise between, on the one hand, the need to cover a
large amplitude and to reduce the instrument drifts, and, on the other
hand, the need to distinguish point sources detection from parasitic
"spikes", and respect both the detectors and the stellar sensor
response times.  The total gondola weight will be of the order of $\rm
1100\,kg$.

\begin{figure}[htb]
\centering
\includegraphics[width=15cm]{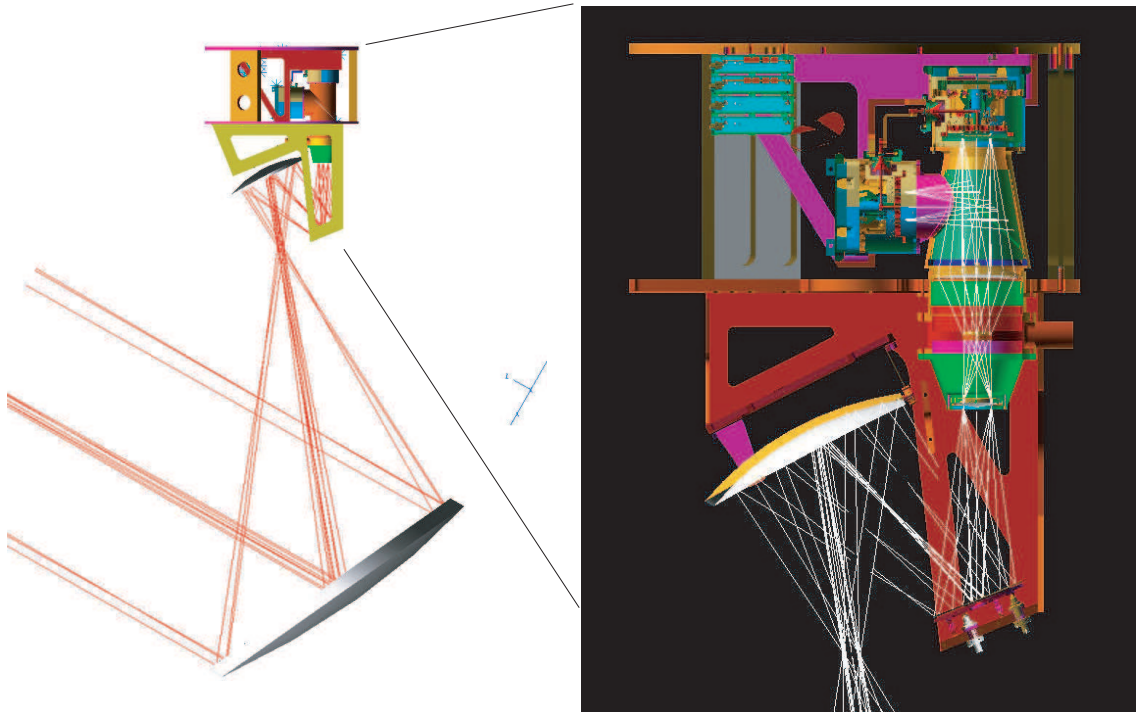}
\caption{\footnotesize Schematic view of the \Pilot Optics and Photometer. Left:
Full view showing the off-axis primary mirror (bottom) and the
cryostat (top). The incoming beam is focalized by the primary on the 
entrance window of the cryostat. Right: blow-up view of the cold
optics inside the cryostat. The two orthogonal polarization
components are split by a tilted polarizer towards the
transmission (top) and reflection (left) bolometer array housings. 
\label{figoptics}}
\end{figure}

The optics of \Pilot is designed to provide an instantaneous field of
view of $1\degr \times 0.76 \degr $, with an equivalent focal distance of
$\rm 1790\,mm$ and a F/2.6 numerical aperture.  The angular resolution of
the instrument ($< 3.5'$) has been chosen as a compromise between the
need of having a resolution similar or better than that of the IRAS
satellite and the aim of surveying large portions of the sky per
flight, with a telescope size limited by weight constraints.  The
optics consists of an off-axis Gregorian telescope with a diameter of
$\rm \simeq 800\,mm$ and a reimaging refractive objective (see
Fig.\,\ref{figoptics}).  The telescope is composed of an off-axis
paraboloid primary mirror and a off-axis ellipsoid secondary mirror,
both made out of Aluminium.  The combination is equivalent to
an on-axis parabolic system (Mizuguchi-Dragone condition) in order to
minimize depolarization effects.  The reimaging refractive objective
consist of two lenses acting as a telecentric objective, reimaging the
focus of the telescope onto the detectors. A Lyot-Stop is located
between the lenses at a pupil plane which is conjugated to the primary mirror.
The polarization will be measured using a rotating half-wave plate
located next to the Lyot-stop and a fixed polarizer in front of the
detectors.  The fixed polarizer will be tilted to about $45\degr$ in
order to reflect one polarization component on one bolometer housing
(reflection array) and to transmit the other polarization component
on a second bolometer housing (transmission array).
This optical configuration is optimized to have good optical
performances and to minimize straylight, both internal and external to
the cryostat.

In order to reduce the background level on the bolometers, all optical
elements except the primary mirror will be located inside a large liquid
He cryostat, cooled down to 2 K using a pumped He bath.  The detectors
will be further cooled down to 0.3 K using an He$^{3}$ closed cycle
fridge mounted on the 2 K plate. The cryostat will also allow the
thermalization on the different cryogenic stages of more than 150 wires
necessary to carry the scientific and house-keeping signals on their
way out toward the warm electronics.  Optical filters will be mounted
both next to the entrance window of the cryostat and close to each
detector housing, where the bandpass selection of each photometric
channel will be performed, so that half of each (reflection and
transmission) detector housing operates in one of the two photometric
channels (240 and $550\mic$).

\begin{figure}[htb]
\centering
\includegraphics[width=1.0\linewidth]{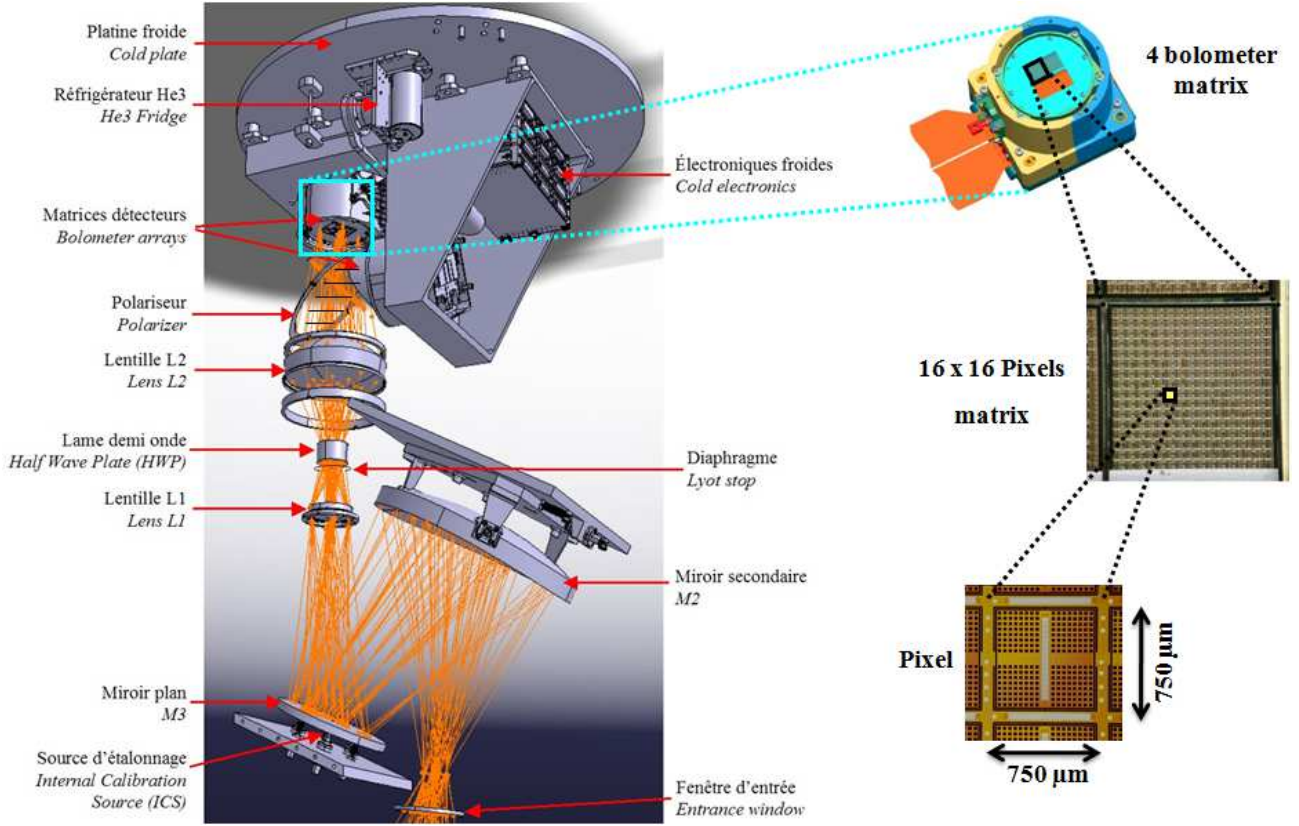}
\caption{
\footnotesize Schematic view of the cryostat showing the location of
the cold optical elements, Internal Calibration Source (ICS) and
bolometer mechanical housing for both photometric channels.  The
inner structure of one of the bolometer housing is shown. The closeup view
reveals the structure of several bolometers of the array.
\label{fig:detectors}}
\end{figure}


We will use the filled bolometer arrays which have been developed by
CEA/LETI for the PACS instrument on board the Herschel satellite (see
Fig.\,\ref{fig:detectors}).  These incorporate fully multiplexed
readout at 300 mK. They are produced as filled arrays of 16*18
detectors.  They are then assembled into a mechanical housing which
ensures cooling at 300 mK and includes the multiplexing and
amplification circuits.  We will use one such housing per polarization
component, each channel being equipped with 4 matrices or 1024
individual detectors.  The Noise Equivalent Power (NEP) of each
bolometer is of the order of a few $10^{-16}\,W/\sqrt{Hz}$, including
noise from the readout electronics.  The readout speed of this type of
array is however much slower than for individual bolometers such as
those which equip the \Archeops and Planck experiments.

Since the polarization measurements are derived from differences
between bolometer signals, either instantaneous of different detectors,
or of the same detector at different instants, very accurate
intercalibration of the bolometer signal at all timescales is
mandatory.  This will be achieved though the use of an Internal
Calibration Source (ICS).  We will use the spare model of the SPIRE
internal calibration source developed for the SPIRE instrument on board
Herschel. This source, passed through a hole of the flat mirror
 within the cryostat, will shine light through the lenses
 (see Fig.\,\ref{fig:detectors}) so as to fully illuminate
both arrays with a highly reproducible illumination pattern.  It will
be used between sky scans to calibrate the variations of the response
flat-field of the detectors.

\subsection{Polarization measurement and intercalibrations}

The polarization will be measured using a combination of a rotating
half-wave plate and the fixed analyzer grid positioned in front of the
detector housings.  This is a classical design in which the polarized
part of the incident light is phase-shifted by the wave-plate and
selectively transmitted by the analyzer.  It produces a modulation of
the polarized signal at twice the rotation frequency of the plate.
Such systems are often used at rotation frequencies of a few Hertz
followed by in-phase analysis of the bolometer signal.  However, such a
use has one main disadvantage: Given that the instrumental background
can easily be $10^5$ times higher than the polarized signal to be
measured, any fluctuation of the background introduced by the rotating
plate, such as anisotropies of the plate transmission, can easily
produce a signal in excess of the astrophysical polarized signal and be
mistaken for polarized signal.  For mapping experiments where scanning
of the sky is obtained by constant drifting, the polarization
modulation ends up being at a frequency close to that corresponding to
the sky structures to be measured.  If well adapted to fast readout
single bolometers, this technique is not necessarily the most optimal
for slower bolometer arrays.  For \Pilot, we plan to use a stepping
wave-plate, the angular position of which (at least 2 are necessary to measure
polarization) will be changed at the end of each scan on the sky.  This
measurement method is similar to what is achieved using Polarization
Sensitive Bolometers (PSBs) or Orto-Mode Transducers (OMT) pairs.  The
$I$ and $Q$ Stokes parameters are measured at a given time $t$,
through a differential measurement and $I$ and $U$
parameters are then measured at a later time $t'$, the main difference
being that, in our case, the time difference between the two
measurements is the time separation between successive scans.  This is
equivalent to modulating the polarized signal at a very low frequency,
which is well below those associated to the structures to be detected
in the maps.  We therefore expect that the background signal should be
slightly different between the different positions of the plate, but
this will contribute a constant offset between individual scans, which
can be removed during data processing and will not affect the
structures observed along the scan.  A second advantage is that,
individual scans being obtained at a constant direction of the
analyzer, redundancy in the scans can be used to intercalibrate the
response of individual bolometers (response flat-field) at any time.
This will be used to monitor the possible variations of the response
flat-field between successive calibration sequences on the ICS.
However, such low frequency modulation does not necessarily freeze the
low frequency drifts of the bolometer noise (so called 1/f noise) or
fluctuations of the residual atmospheric signal with sufficient
accuracy.  Results of numerical simulations
for these contributions indicate that this will not affect our ability
to recover the polarization intensity, partly because the 1/f noise of
the bolometers used is low.


\section{Ground Tests}

The aim of the \Pilot ground tests was to validate the instrument
behavior and to optimize its performances. These tests were realized
in a clean room at the CNES facilities in Toulouse, France. They are
described in the next subsections. Since the background falling on the
detectors is much stronger in the ground configuration than at ceiling
altitude, and the detectors are optimised for a moderate background,
of the order of 5 pW/pix, 
we inserted an additional filter in the filter stack, which attenuates a large fraction of the radiation by a factor of 20.
The spectral transmission of the attenuator was measured with the FTS and set according to 
the predictions of a photometric model of the instrument taking into
account external and internal sources of emission, as well as the
transmission of the various optical filters, as measured at the
subsystem level.

\begin{figure}[htb]
\centering
\includegraphics[width=0.95\linewidth]{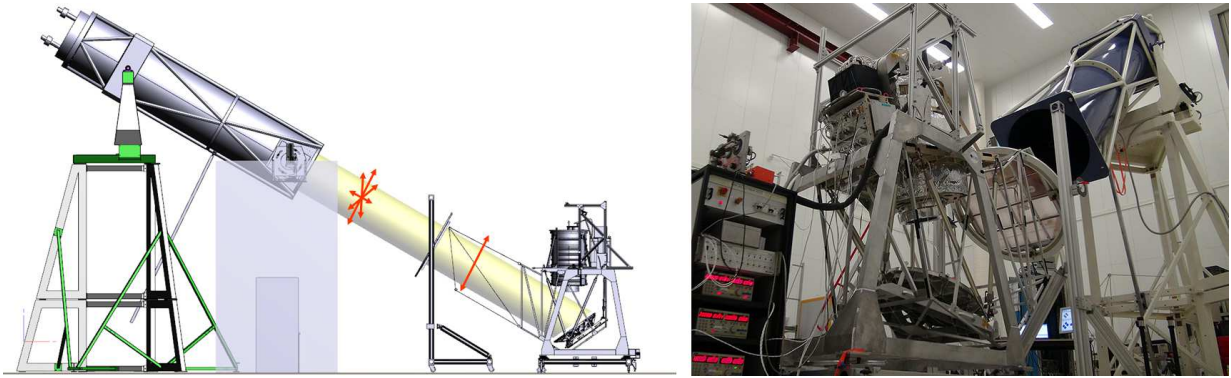}
\caption{Optical bench used to characterize polarization
properties of the instrument, showing from left to right, the IRAP
$\rm \phi=1\,m$ collimator, the large format \Archeops polarizer and
the \Pilot pointed load. The polarizer can be removed for intensity
only measurements.
\label{fig:conf_test_ARPOB}}
\end{figure}

For these tests, the \Pilot pointed load was aligned with the IRAP
submillimeter bench (see the experimental setup in Fig.\,\ref{fig:conf_test_ARPOB}). 
The optical bench is composed of a high
intensity lamp (Mercury plasma lamp) at the focus of a $\rm \phi=1\,m$
diameter Newton telescope with a focal length of 5340 mm, delivering a $1 m$ diameter collimated
beam, and acting as a point source at infinite distance. The source is
modulated by a rotating chopper. The source is mounted on a motorized
2D moving plate and can be moved at the focus of the system, so that
it can be displaced over a limited range without moving the telescope.


\subsection{OPTICAL PERFORMANCES}

\subsubsection{FOCUS}

The requirement on the overall optical quality of the \Pilot
instrument is that the alignment between the primary mirror and the
photometer requires accuracy of the order of $\pm 300 \mu m$ for the
focus, and a maximum mismatch of $0.06 \degr $ 
between optical axes of their optical axes. 
The requirement must be satisfied at all time
during the observations at ceiling altitude, while the elevation of
the pointed load is changed and the temperature of the structure
evolves. This is actually one of the most stringent requirements for
the experiment.

Fulfilling this requirement has implied a series of actions: First, 
precise characterization of the mechanical and optical properties of
the primary mirror alone and of the photometer alone have been
carried. The mirror characterization \cite{Engel2013} was performed
with the submillimeter test bench shown in
Fig.\,\ref{fig:conf_test_ARPOB} with no polarizer.
The photometer optical characteristics were determined using a ZEMAX
simulation and $3D$ measurements.  Second, specific means and an
efficient procedure to align the two subsystems, both in the
laboratory and at the launch site, are needed. To perform this
alignment, the primary mirror is mounted on an hexapod, which allow
six degrees of freedom for the adjustment while maintaining the
stiffness of the system. A numerical model of this mechanism allows a
fast convergence of the optical alignment. The relative position of
the two subsystems is checked by measuring, using a laser tracker
system, of a series of optical reference balls associated with the
mirror and the photometer.  Third, despite the fact that the whole
structure connecting the photometer and M1, and M1 itself are made of
the same material (aluminium), which limits differential dilatation,
the optical system is off-axis and no perfect compensation is
possible. In order to control the optical alignment during the
flight, a thermal model of the instrument has been developed. Before
the launch, the use of this model knowing the flight conditions, will
allow to correct for the expected displacement, compensating to first
order the effect of the thermal expansion of the mechanical structure
expected at ceiling altitude.  Fourth, deformation of the holding
structure due to gravity at the various elevations of the pointed load
has been modelled (using a finite elements analysis based on
Nastram). The deformations are expected to be somewhat lower than
thermal deformations. They have also been measured during tests.

In order to determine the best optical alignment of M1 with respect
to the photometer, we have performed a series of defocusing along
three orthogonal axes, around the best theoretical focus supplied by the
Zemax model. The Z defocusing direction corresponds to the optical
axis of the photometer while X and Y are orthogonal to it.  To
estimate the impact of defocusing on the optical performance, several
parts of the focal plane have been explored with the collimator for
each defocus.
The source PSF was fitted in the obtained images, in order to measure the
changes in the Full Width Half Maximum (FWHM) and encircled energy as
a function of defocusing distance.  The best focus position was
determined as the position of minimum PSF FWHM.

\begin{figure}[!htb]
\centering
\includegraphics[width=5.1cm]{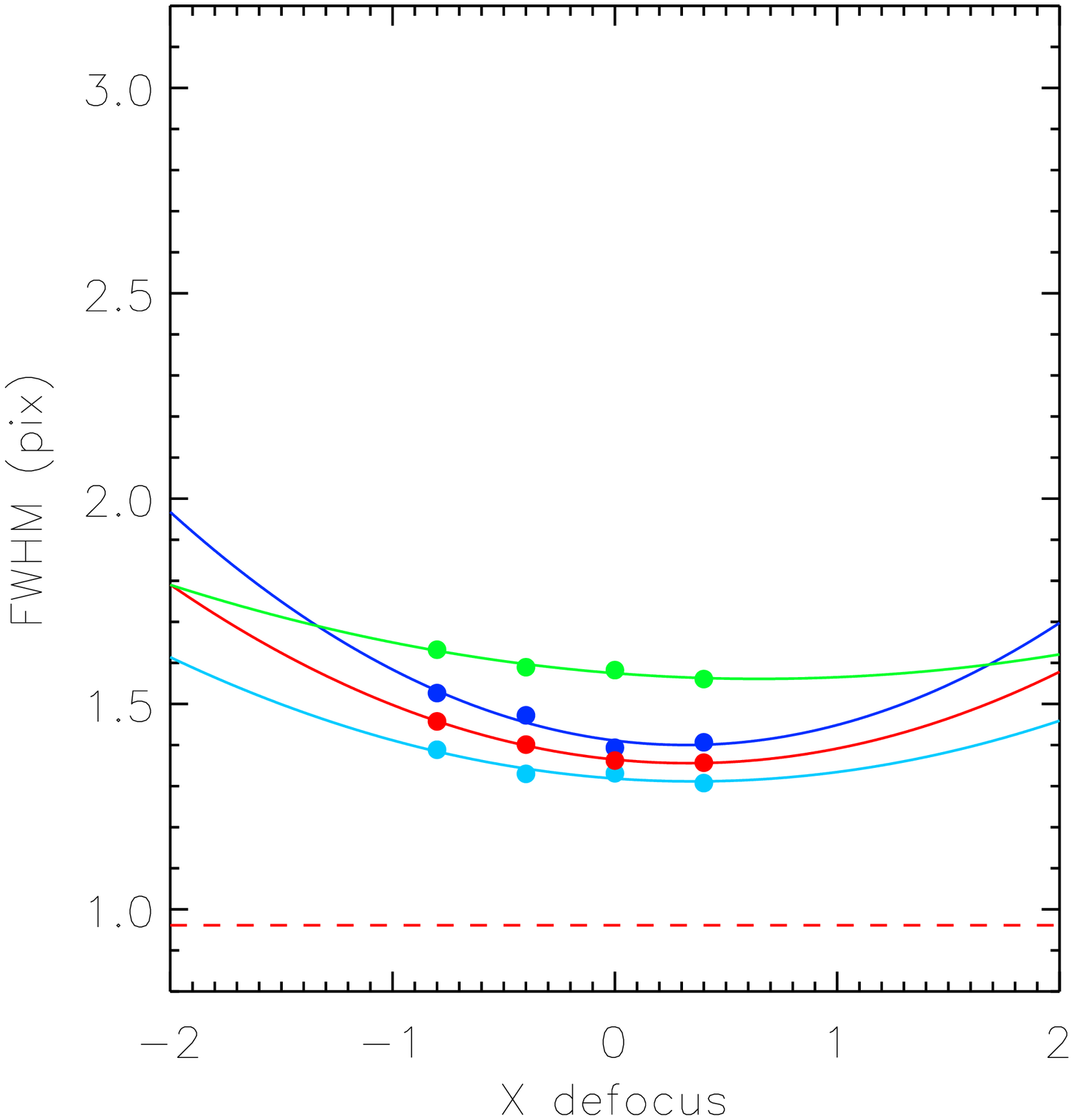}
\includegraphics[width=5.1cm]{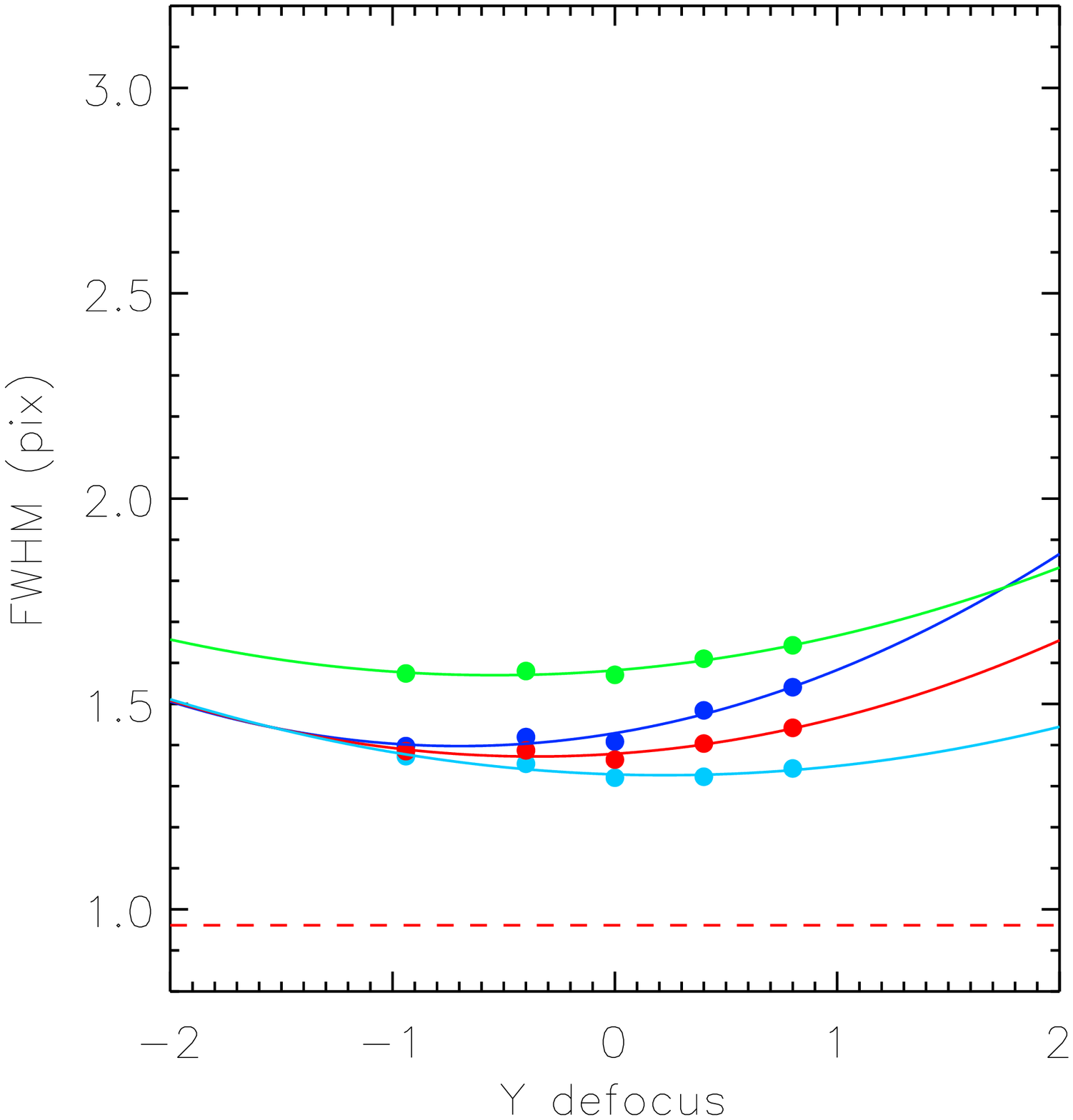}
\includegraphics[width=5.1cm]{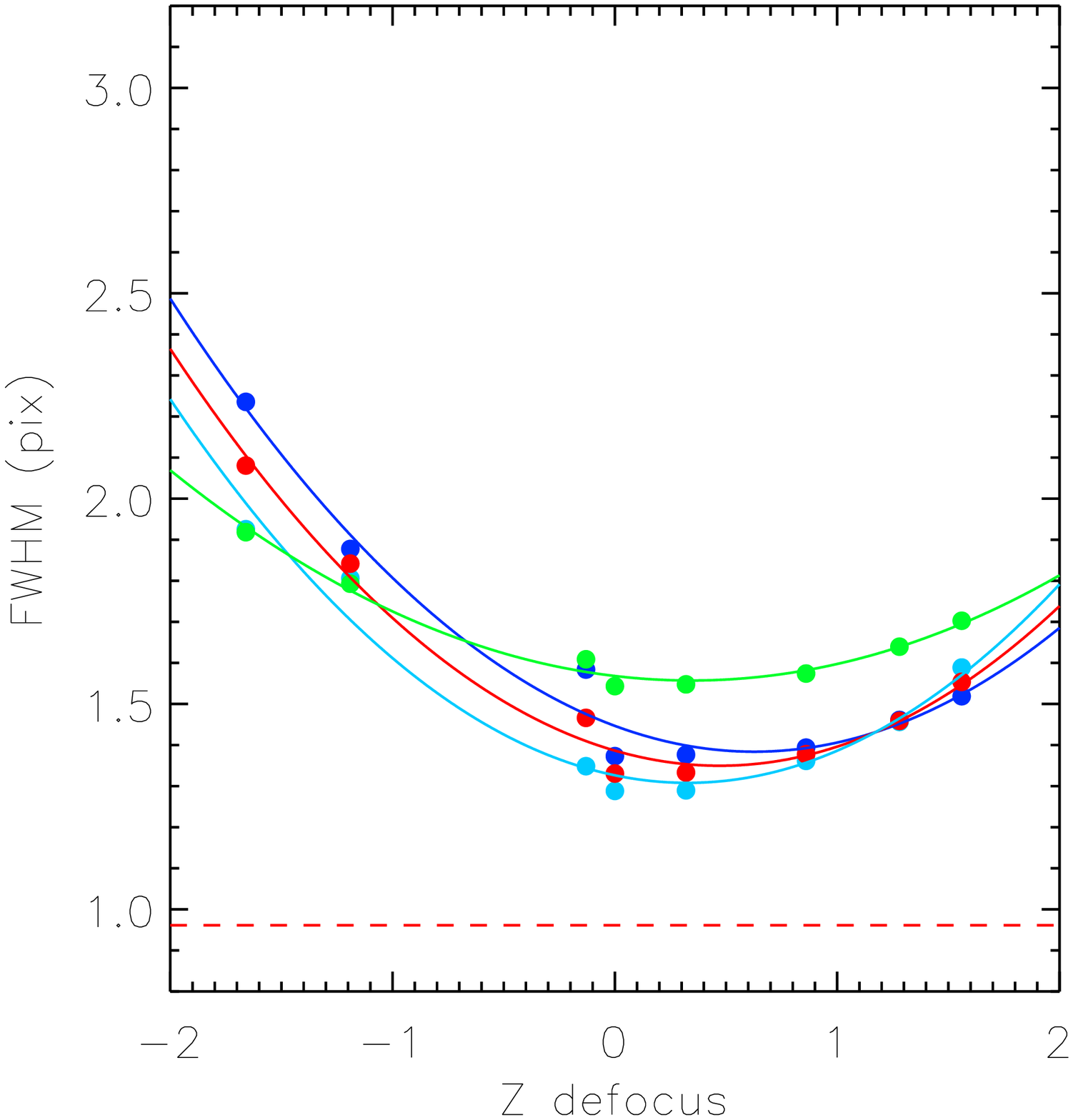}
\caption{Measured FWHM of the collimator source, as a function of the defocus
distance along the X (right), Y (center) and Z (left) on \Pilot array
\#2. FWHM values are in array pixel units.  The horizontal dashed
lines show the diffraction limit.  Blue lines show the FWHM along the
X axis and Cyan lines show the FWHM along the X axis, the red lines
show averaged FWHM of X and Y axis and green lines show the FWHM
values derived from a circular profile of the PSF.
\label{fig:defocus}}
\end{figure}

Figure \ref{fig:defocus} shows an
example of the FWHM of the observed source on an array at some defocus
positions, the defocus distance measured from the initial position.

\subsubsection{POINT SPREAD FUNCTION}
\begin{figure}[!htb]
\centering
\includegraphics[width=8cm]{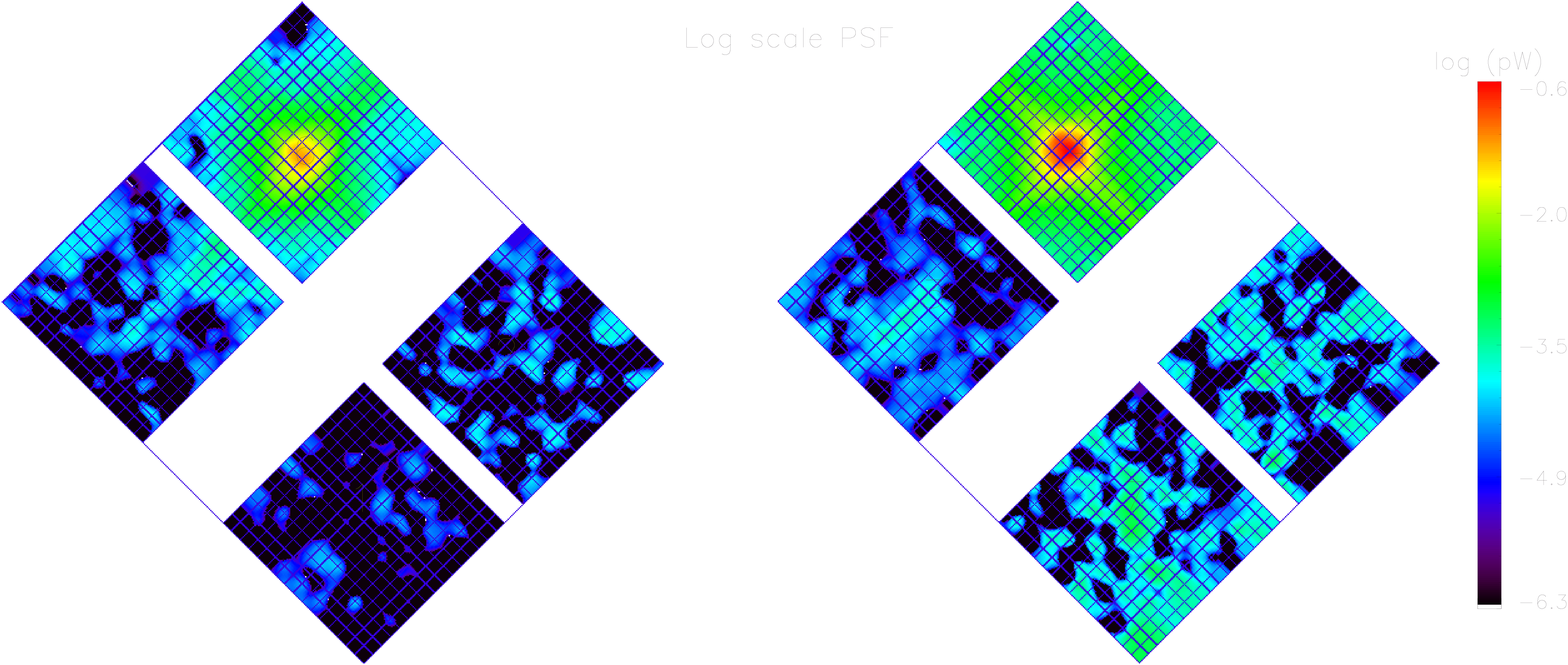}
\includegraphics[width=8cm]{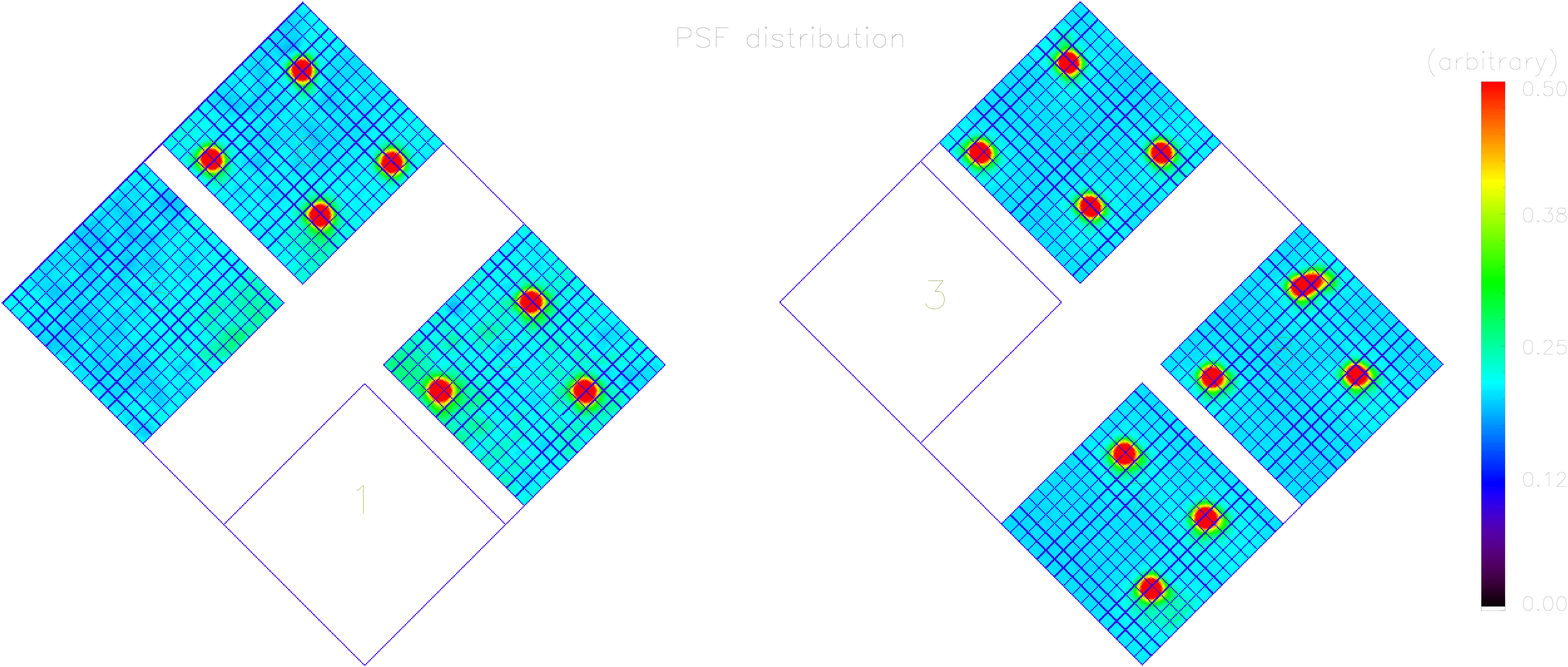}
\caption{
Left: Image of a point source on the \Pilot focal plane. The left and right
part of the figure show the Transmission and Reflection arrays
respectively, as projected on the sky with the elevation axis
increasing upward and the cross-elevation axis being horizontal. The
intensity is in log scale. Right: Distribution of PSF shapes
on the \Pilot focal plane for the X,Y,Z=0,0,0 defocus position.
\label{fig:PSF_log}}
\end{figure}

The collimator is used to simulate a point source at infinite distance which sky position is accurately known through laser tracker measurements.
An example of the observed \Pilot Point Spread Functions (PSF) is represented in Fig.\,\ref{fig:PSF_log} for a transmission and a reflection array.
The origin of the position offset is the peak intensity position of the PSFs along the x and y axis.
The PSFs are axial symmetries and along the x axe and the y axe are also in symmetry.

\begin{figure}[!htb]
\centering
\includegraphics[width=4.5cm]{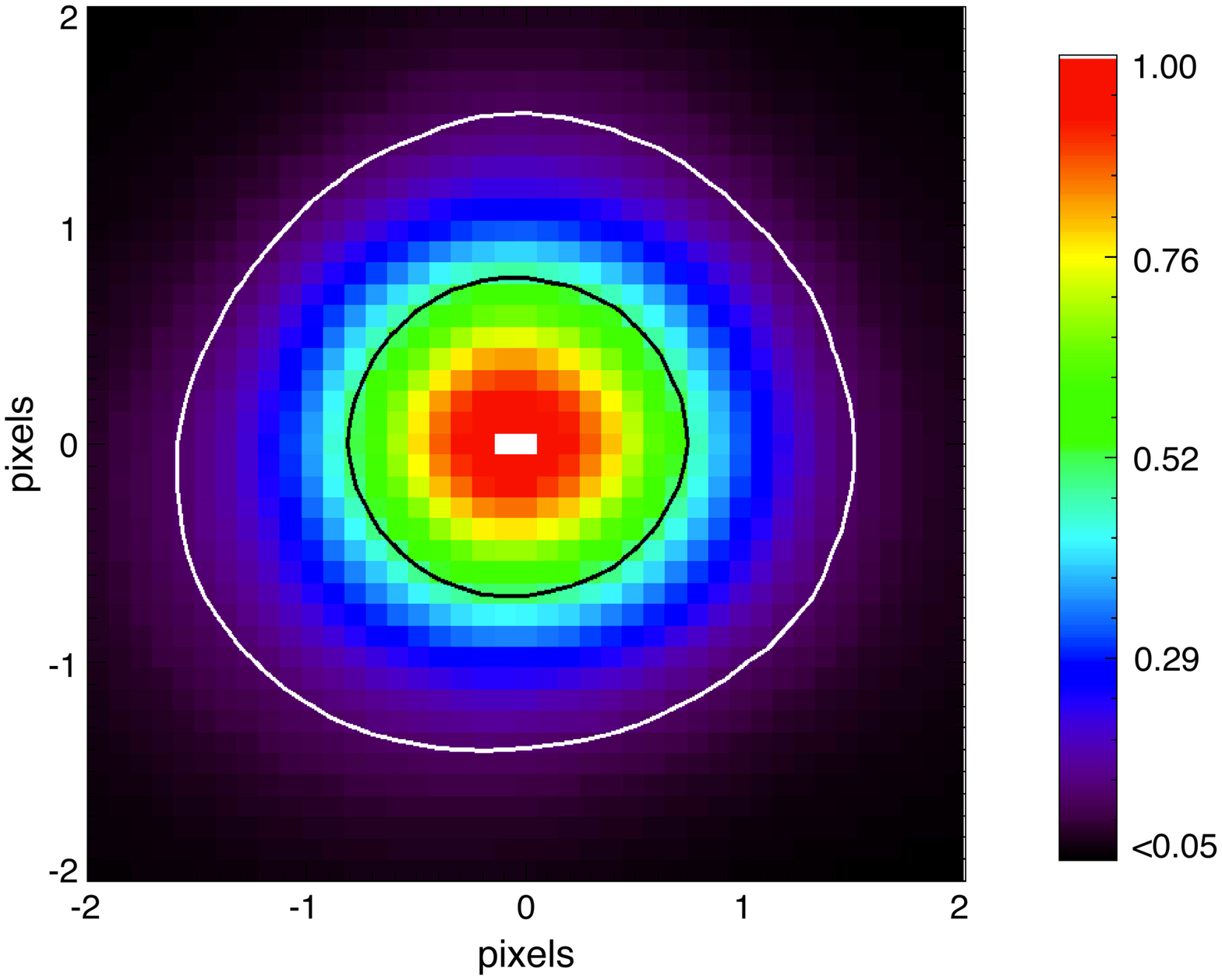}
\includegraphics[width=5.5cm]{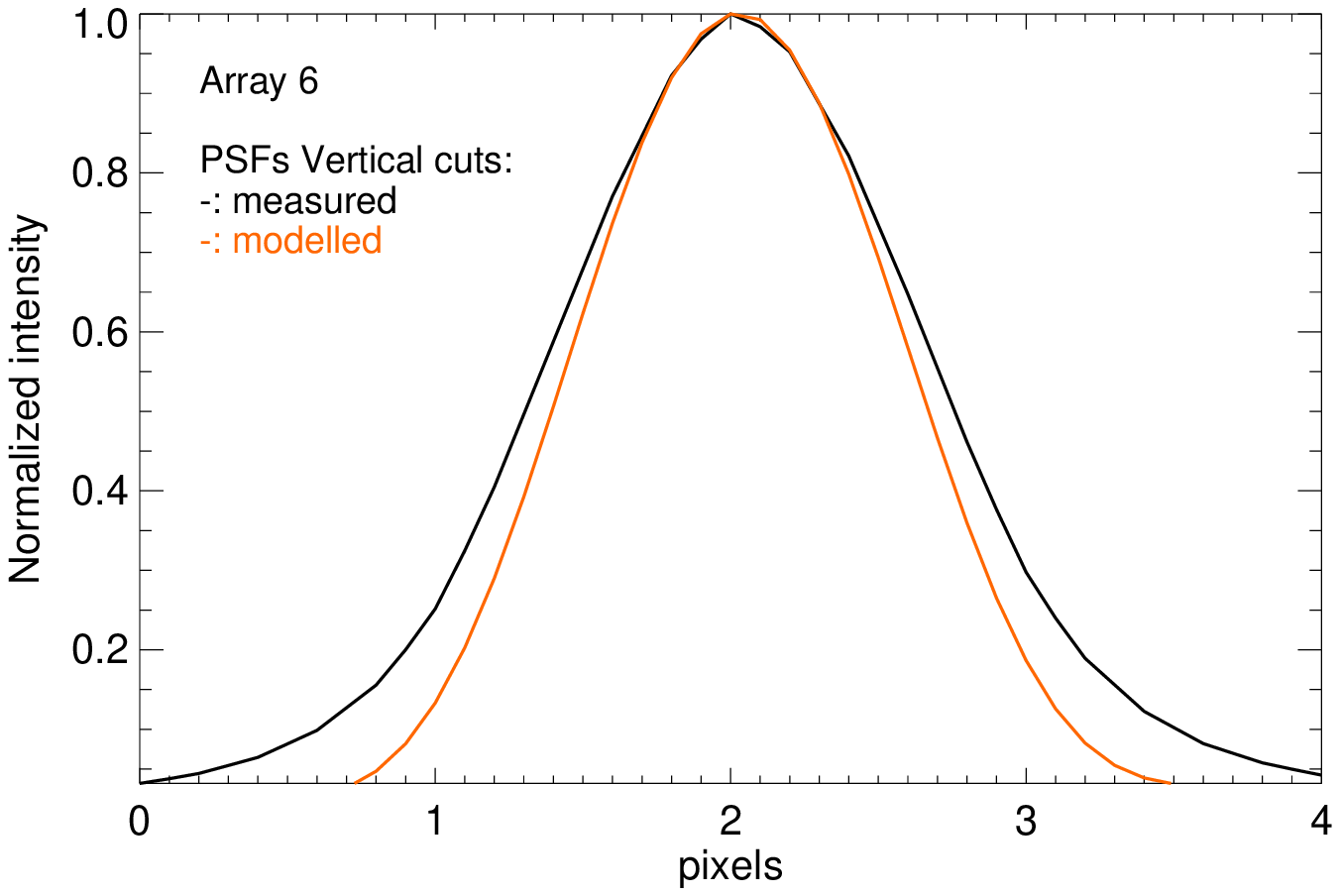}
\includegraphics[width=5.5cm]{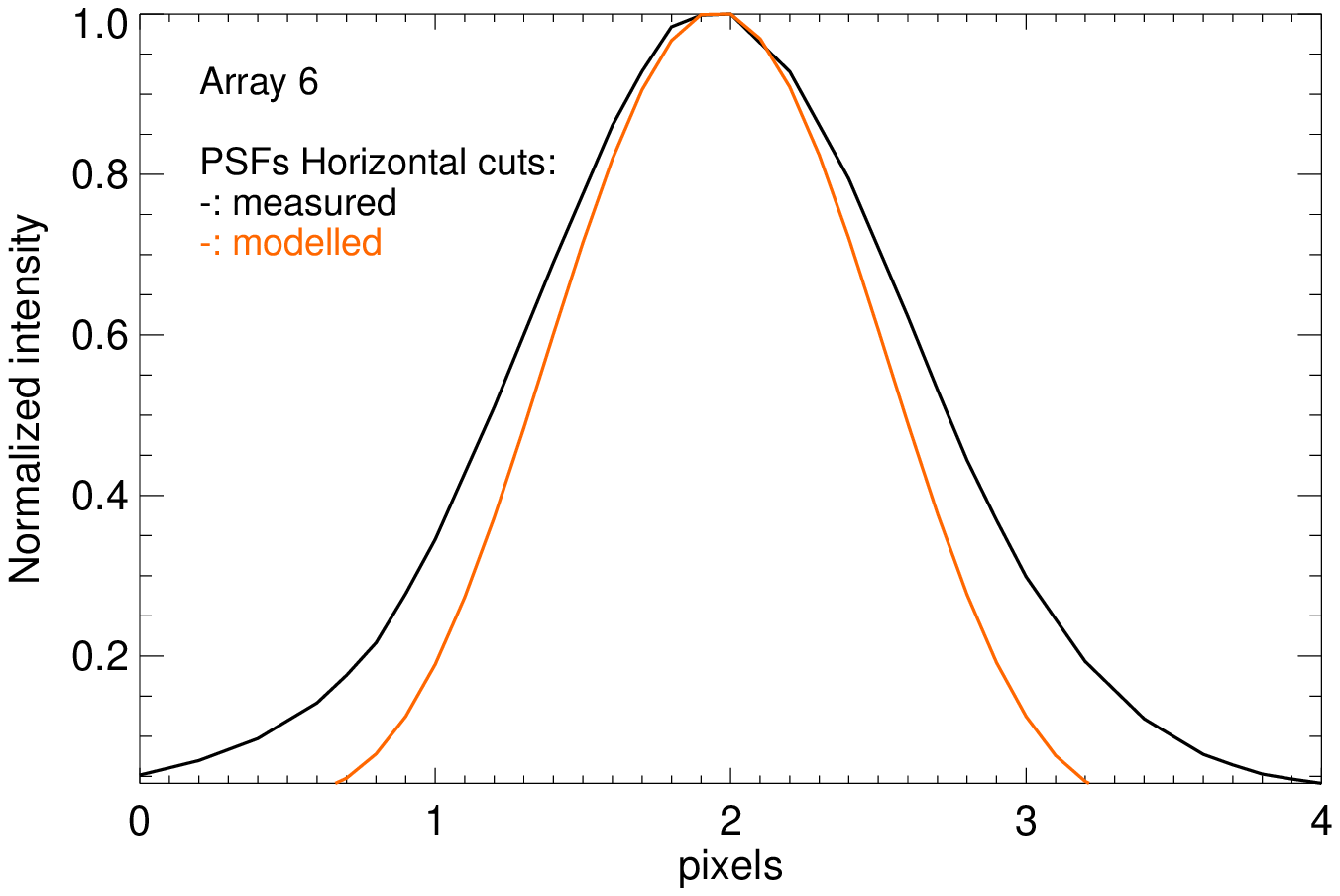}
\caption{
Left: Image of the average PSF obtained from a
$12 \times 12 $ microscaning pattern over a region of $\pm2$ pixels around
pixel $(3,7)$ of array \#6. Contour levels are at 1.0, 0.5 and
0.1. Center and Middle:
The vertical (center) and horizontal (middle) normalized cuts across the
measured PSF (black) and the modelled PSF (red).
\label{fig:PSF_data_image}}
\end{figure}

Figure\,\ref{fig:PSF_data_image} shows the average PSF obtained from a
$12 \times 12 $ microscaning pattern over a region of $\pm2$ pixels around
pixel $(3,7)$ of array \#6. Figure\,\ref{fig:PSF_data_image} shows the
comparison between the above PSF and the modelled PSF.  The observed
PSF shows reasonably good agreement with the modelled PSF.

\subsection{FOCAL PLAN GEOMETRY}

The aim of this test is to establish the precise correspondance
between positions in the focal plane and directions in the sky.  For
this test, the sub-millimeter bench shown in
Fig.\,\ref{fig:conf_test_ARPOB} was used without polarizer. The
collimator was calibrated
using a theodolit to provide the correspondence between source
displacement at the focus of the collimator $(X_S,Y_S)$ and the elevation and
cross-elevation offsets of the source $(\Delta_{el};\Delta_{cel})$
direction.

We measure the \Pilot focal plan geometry using the collimator source
positionned at the centre and the 4 corners of each array,
at several elevations of the collimator (23, 30, 45 and 49$\degr$).
Each $(X_S,Y_S)$ position of the source and the associated elevation and cross-elevation
angles with respect to the center of the instrument focal plane have
been measured, are dated
by the on board computer and saved with the data.
For each position of the source, we compute the average PSF image from
the average ON and OFF images according to the
collimator shopper position. Each PSF is fitted using a 2D Gauss fitting
procedure.

We fit the following system:
\begin{equation}
\begin{pmatrix} \Delta_{cel}-\delta_{cel}^0 \\ \Delta_{el}-\delta_{el}^0 \end{pmatrix}=
\begin{pmatrix} \delta_x \\ \delta_y \end{pmatrix}
\begin{pmatrix}  cos(\alpha) & sin(\alpha) \\-sin(\alpha) &  cos(\alpha) \end{pmatrix}
\begin{pmatrix} x-x_0 \\ y-y_0 \end{pmatrix}
\label{equ:fpgeom_fit}
\end{equation}
where $x$ and $y$ are the measured pixel position of the peak of the
PSFs along the array directions, $x_0$ and $y_0$ are the pixel
position of the rotation center,
$\alpha$ is the rotation angle,
and $\delta_x$ and $\delta_y$ are the pixel scale along the
array directions. $\Delta_{cel}$ and $\Delta_{el}$ are
the known cross-elevation and elevation offsets of the collimator source compared
to a reference position (API). $\delta_{cel}^0$ and
$\delta_{el}^0$ are the cross-elevation and elevation offset between
the API and the reference pixel ($x_0$,$y_0$). 
We let $\delta_{cel}$ and $\delta_{el}$, $\alpha$, $\delta_{cel}^0$ and
$\delta_{el}^0$ be free parameters. $x_0$, $y_0$ are set at the center
of each array.

The average focal plane geometry parameter values derived for each arrays
fitting Eq.\,\ref{equ:fpgeom_fit}
are given in Tab.\,\ref{tab:fp_geom}. Uncertainties given in the table are computed from
the standard deviation of free parameters for the various elevations.

\begin{table*}
\caption[ ]{\label{tab:fp_geom} Elevation-averaged focal plane geometry
\small
  parameter values obtained. The vertical line separates arrays from
  the transmission (TRANS) and reflexion (REFL) array respectively.}
\begin{flushleft}
\begin{tabular}{llllllll}
\hline
\hline
Array & $\delta_x$ & $\delta_y$ & $\alpha$ & $x_0$ & $y_0$ & $\delta_{cel}^0$ & $\delta_{el}^0$ \\
         &  '/pix &  '/pix & degree & pix & pix &  pix & pix \\
\hline
    2   &  -1.476$\pm$0.016  &  1.480$\pm$0.009  &  -46.481$\pm$0.452  &  7.5$\pm$0.0  &  7.5$\pm$0.0  &  22.405$\pm$0.371  &  -5.255$\pm$0.302\\
    6   &  -1.530$\pm$0.006  &  1.416$\pm$0.006  &  -43.470$\pm$0.101  &  7.5$\pm$0.0  &  7.5$\pm$0.0  &  -4.851$\pm$0.379  &  20.557$\pm$1.234\\
    5   &  -1.488$\pm$0.003  &  1.476$\pm$0.002  &  -42.219$\pm$0.174  &  7.5$\pm$0.0  &  7.5$\pm$0.0  &  -22.777$\pm$0.327  &  2.054$\pm$1.162\\
\hline
    4   &  1.529$\pm$0.004  &  1.418$\pm$0.010  &  44.414$\pm$0.264  &  7.5$\pm$0.0  &  7.5$\pm$0.0  &  -4.688$\pm$0.345  &  20.630$\pm$1.252\\
    8   &  1.478$\pm$0.001  &  1.483$\pm$0.003  &  47.369$\pm$0.405  &  7.5$\pm$0.0  &  7.5$\pm$0.0  &  22.147$\pm$0.344  &  -5.965$\pm$0.314\\
    7   &  1.513$\pm$0.005  &  1.498$\pm$0.006  &  46.183$\pm$0.260  &  7.5$\pm$0.0  &  7.5$\pm$0.0  &  3.406$\pm$0.435  &  -24.456$\pm$1.312\\
\hline
\end{tabular}
\end{flushleft}
\end{table*}




\subsection{BACKGROUND AND OFFSETS}

The aim of these tests is to measure the amplitude and spatial shape
of the background. These measurements are also used to derive the
electronic offsets of the detectors and to measure the instrumental
polarization. For these tests, the experience has been placed under a
controlled Nitrogen atmosphere (relative humidity $ < 2 \%$) in order
to diminish absorption between the emission source and the instrument,
which is mostly contributed by water in the atmosphere.

\noindent Three configurations were used for these tests: First, a $1
m^2$ eccosorb sheet, acting as a room temperature black body, was
placed in front of the whole instrument, covering the whole the whole
instrument field of view.  Second, the \Pilot primary mirror was
removed and a tank containing eccosorb maintained either at room
temperature or at Liquid Nitrogen ($77 K$) temperature was inserted in
from of the photometer, aginst covering the whole FOV of the instrument.



The measurement $m_T$ and $m_R$ on the Transmission and Reflexion
arrays of PILOT are related to the radiation intensity $I$,
polarization fraction $p$ and polarization angle $\psi$ through
\begin{eqnarray}
m_T & = & \frac{1}{2} R_{xy}^T \times T_{xy}^T I\times(1+p \cos{2 \psi} \cos{4\omega} + p \sin{2 \psi} \sin{4\omega})+O^R_{xy} \label{Eq:signalT} \,; \\
m_R  & = & \frac{1}{2} R_{xy}^T \times T_{xy}^T I\times(1-p \cos{2 \psi} \cos{4\omega} - p \sin{2 \psi} \sin{4\omega})+O^T_{xy}. \label{Eq:signalR}
\end{eqnarray}
where $T_{xy}$ is the optics transmission and $R_{xy}$ is the detector
response.
$\omega$ is the angle between the HWP fast axis and the vertical direction. $O^R_{xy}$ and $O^T_{xy}$
are the electronic offsets for Reflection and Transmission arrays respectively.

For each pixel, we fit the signal as a function of HWP angle using
\begin{eqnarray}
m_T & = & A^T_0+A^T_1 \cos{4\omega} + A^T_2 \sin{4\omega} \label{Eq:signalfitT} \,; \\
m_R  & = & A^R_0-A^R_1 \cos{4\omega} - A^R_2 \sin{4\omega} \label{Eq:signalR} \,.
\end{eqnarray}
where the fitted parameters $A_0$, $A_1$,
$A_2$ are related to the polarization parameters $I$, $\polfrac$ and $\polang$ as
\begin{eqnarray}
I & = & 2(A_0-O_{xy})/\mathcal{R}_{xy} \label{Eq:I} \\
\polang & = & \frac{1}{2} atan(A_2,A_1) (+\pi/2 for R array)\\
\polfrac & = & \frac{(A_1^2+A_2^2)^{1/2}}{A_0-O_{xy}/2}.
\end{eqnarray}

Assuming that the background sources at 300 K and 77 K have the same
flat true spatial distribution and that the intensity is
proportional to temperature (the Rayleigh-Jeans approximation for
Black-Bodies), ie $I(T) \propto T$,
\begin{equation}
A_0(300K)-A_0(77K) = \frac{1}{2} R_{xy}T_{xy} (300-77).
\end{equation}
The optical transmission (or the shape of the background illumination on the detector)
is then given by
\begin{equation}
T_{xy} = \frac{2(A_0^{300K}-A_0^{77K})}{(300-77) R_{xy}}. \label{Eq:signalTxy}
\end{equation}
Alternatively, the offsets on both focal planes can be computed using
Eq.\,\ref{Eq:I} applied to one of the background measurements and by
replacing $T_{xy}$ using Eq.\,\ref{Eq:signalTxy}, as
\begin{equation}
 O_{xy} =  A^T_0-\frac{1}{2} T R_{xy} T_{xy} = \frac{77A_0^{300K}-300A_0^{77K}}{(77-300)}.
\end{equation}
Note that the offsets are independent of the system response $\mathcal{R}_{xy}$.



\begin{figure}[htb]
\centering
\includegraphics[width=8cm]{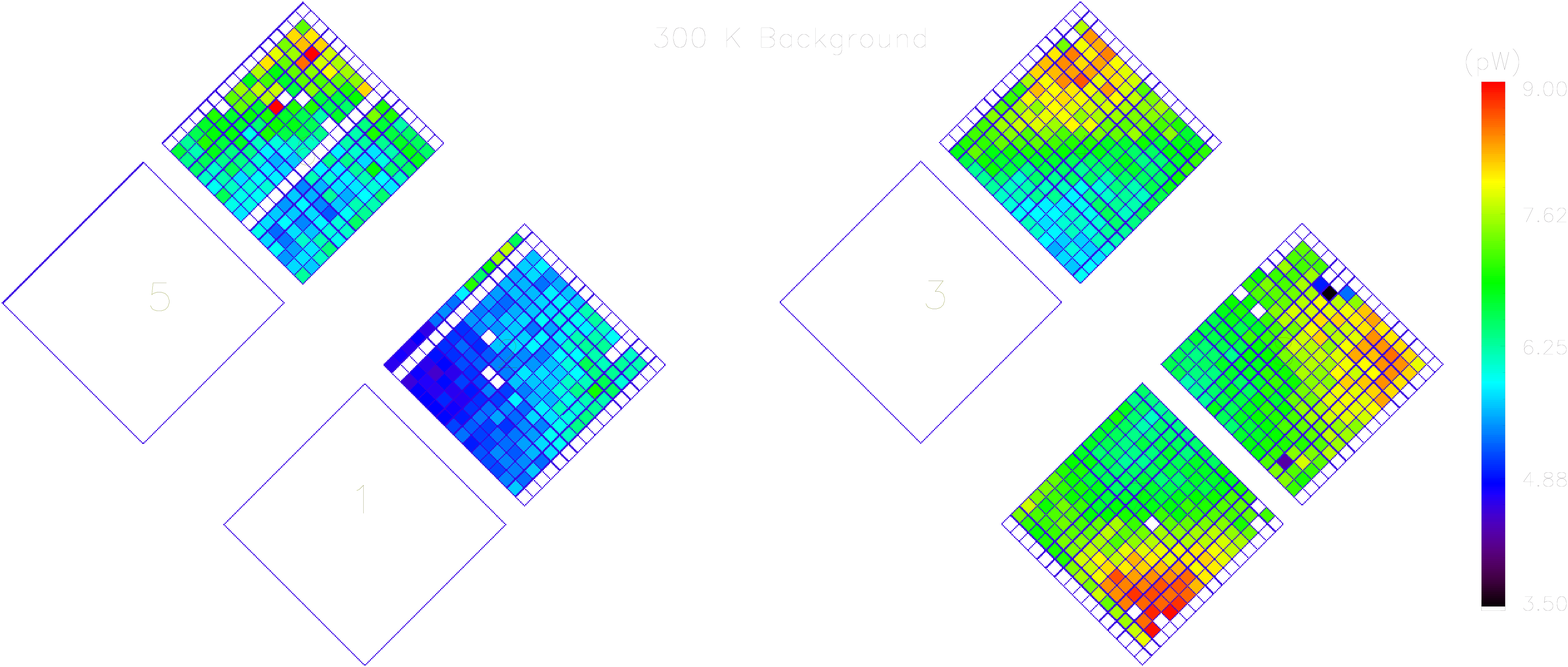}
\includegraphics[width=8cm]{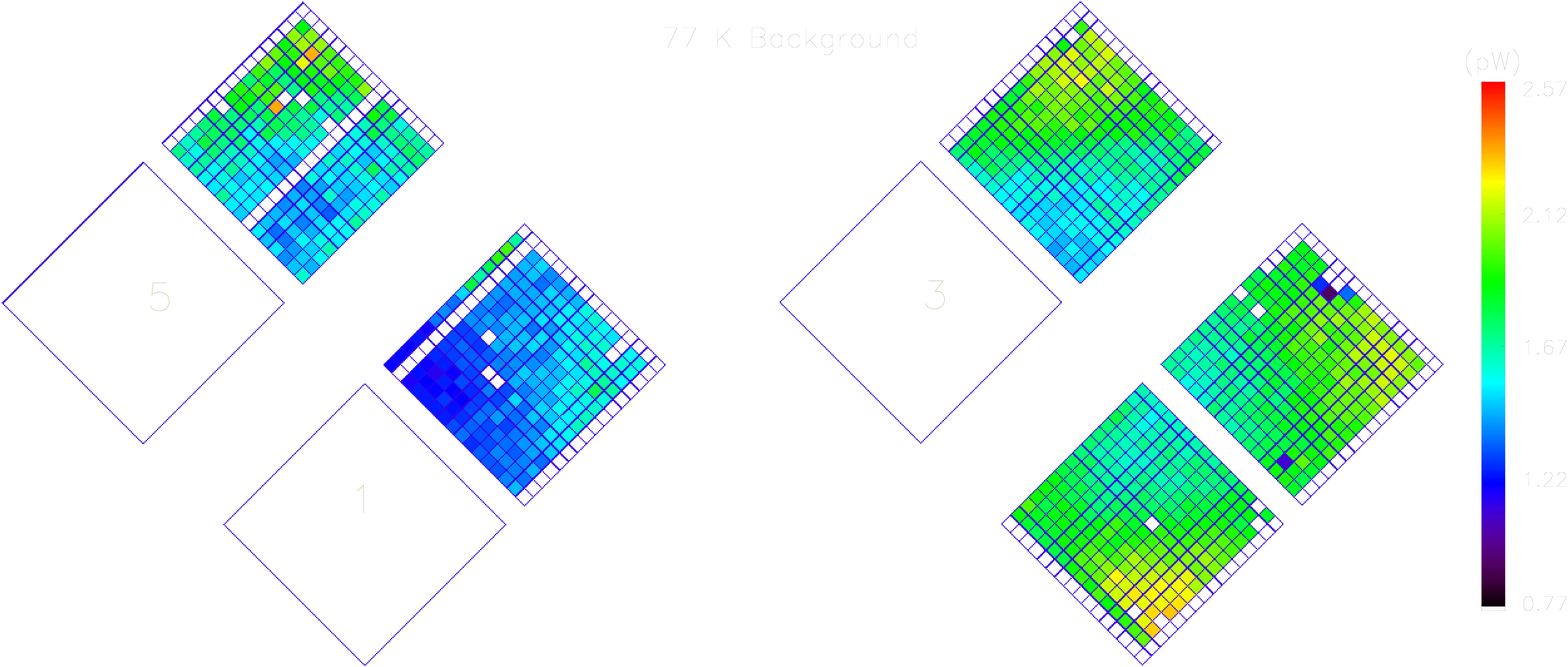}
\caption{
Map of the measured background for the background at 300\,K (left) and
77\,K (right) in the \Pilot focal plane. The color scale is such that
the dynamics in the two images is the same.
\label{fig:back}}
\end{figure}

The obtained background images are shown in Fig.\,\ref{fig:back}.
The color scale is such that the dynamics in the two images is the
same. At 300\,K, the background levels are between 3.5 and 9 pW/pix,
and they are correspondingly lower at 77\,K.
The background images show lower values at the center of the focal
plane, and larger values in outer regions. The variations are
$\simeq$30\,\% and 24\,\% at 300 K and 77\,K respectively. This is likely due to
vignetting by the optics, in particular by lens L2 which has a large curvature.

\subsection{POLARIZATION}

In order to characterize the polarization performances of the
experiment, we use a large format ($\rm 1 \times 1\,m$) polarizer,
developed originally for the \Archeops experiment, hereafter  refered to
as {\archpol}. The {\archpol} polarizer is
composed of an array of $3 \times 3$ frames hosting a large number of
metal wires ($50\mic$ thickness, $100\mic$ separation) and was
initially optimized for millimeter wavelengths. The polarizer
is inserted in the beam of the collimator as shown in
Fig.\,\ref{fig:conf_test_ARPOB}, and can be rotated, in order to
produce a polarized source with any polarization direction. Assuming
that polarization cross talk of the PILOT instrument is negligible,
the polarizer was shown to provide a source polarization fraction
higher than $73\%$ in the \Pilot $240\mic$ band.  The wire direction
of each frame constituting the polarizer was measured with theodolites
and laser-tracker. The median wire direction was measured with respect
to the polariser mechanical structure with a accuracy of about
$2.4'$. A series of optical reference balls associated with the
polarizer structure allow us to measure its orientation with an
accuracy of $1.4''$.  Globally, the direction of linear polarization
is known to an accuracy of about $2.4'$, with respect to the
instrument.

Assuming that the polarization splitting grid inside the has \Pilot
photometer has negligible cross-polarization, it can be shown that the
total power measured by the detectors of the transmission
and reflection arrays can be expressed as
\begin{eqnarray}
m_T &=& 1+\frac{1+2\gamma}{2}k\cos2\alpha + (k\beta\cos2\alpha+\beta)\cos2\omega
+ k\beta\sin2\alpha\sin2\omega \nonumber\\
&+& \frac{1-2\gamma}{2}k\cos2\alpha\cos4\omega +
\frac{1-2\gamma}{2}k\sin2\alpha\sin4\omega \label{eq:mt}\\
m_R &=& 1-\frac{1+2\gamma}{2}k\cos2\alpha + (k\beta\cos2\alpha-\beta)\cos2\omega + k\beta\sin2\alpha\sin2\omega \nonumber\\
&-& \frac{1-2\gamma}{2}k\cos2\alpha\cos4\omega -
\frac{1-2\gamma}{2}k\sin2\alpha\sin4\omega \label{eq:mr}
\end{eqnarray}
where $\beta$ is the differential transmission between the fast and
slow axis of the HWP, $\gamma$ is the phase shift induced by the HWP
thickness.  $\alpha$ is the polarization angle of the incident
radiation given by the angular position of \archpol. $k$ is the
cross-polarization of the \archpol polarizer.  For a perfect HWP,
$\beta=0$ and $\gamma=-0.5$ and Eq.\,\ref{eq:mt} and Eq.\,\ref{eq:mr}
have no term in $2\omega$.


\subsubsection{{\archpol} CHARACTERIZATION}
\label{se:arch_pol}

Another reasonable assumption is that the HWP is very good (which will be
checked in the next section) or at least that its imperfections are small
compared to the likely poor cross-polarization expected from large diameter
\archpol, designed to work at wavelengths larger than 850$\mu
$. We therefore assume in this section that $\beta=0$ and $\gamma=-0.5$. Equations
(\ref{eq:mt}) and (\ref{eq:mr}) therefore reduce to:

\begin{eqnarray}
m_T &=& 1 + k\cos2\alpha\cos4\omega + k\sin2\alpha\sin4\omega \label{eq:mt_arch}\\
m_R &=& 1 - k\cos2\alpha\cos4\omega - k\sin2\alpha\sin4\omega \label{eq:mr_arch}
\end{eqnarray}

\begin{figure}
\begin{center}
\includegraphics[width=4.2cm,trim=4cm 0 0 0]{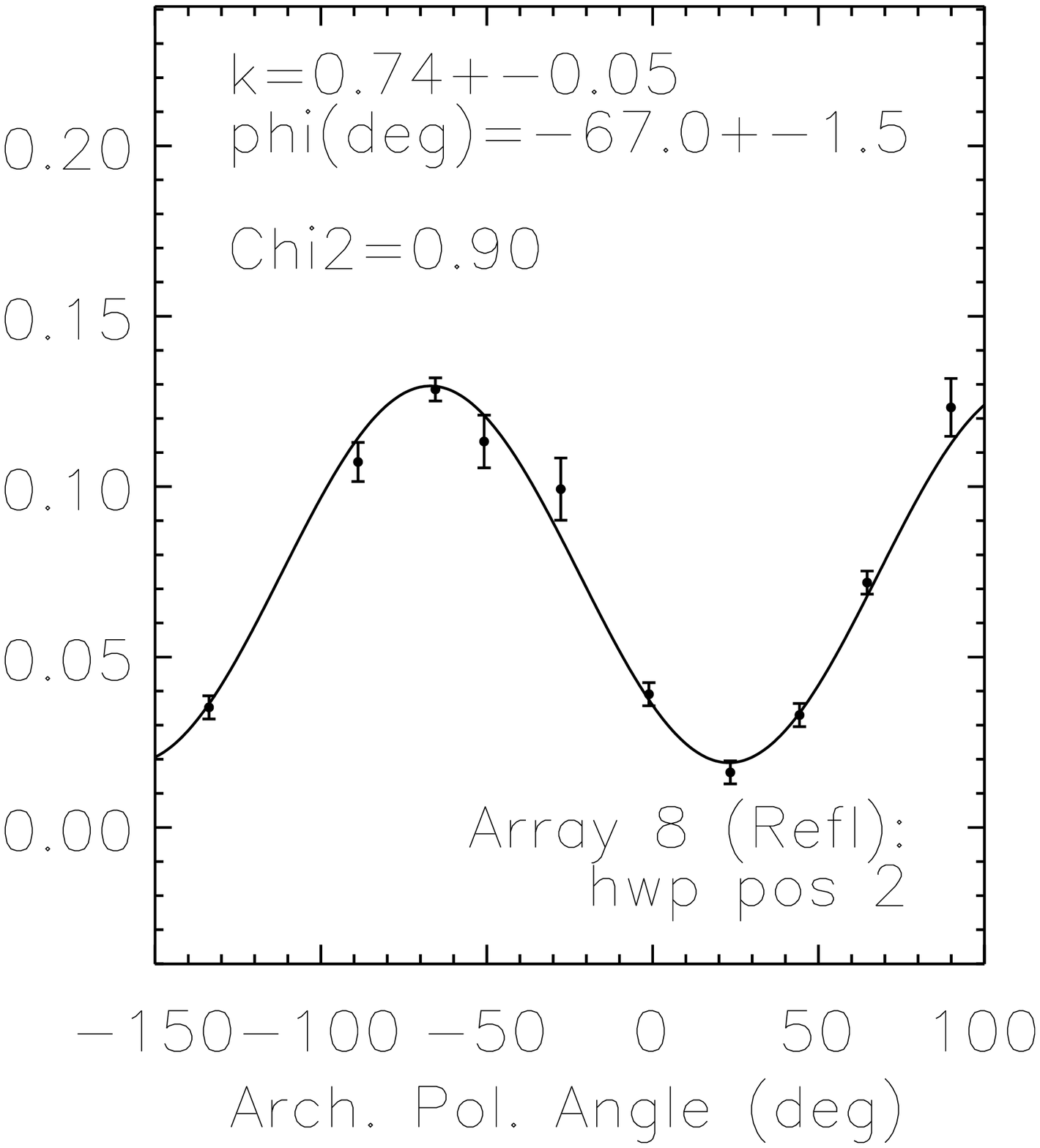}
\includegraphics[width=4.2cm,trim=4cm 0 0 0]{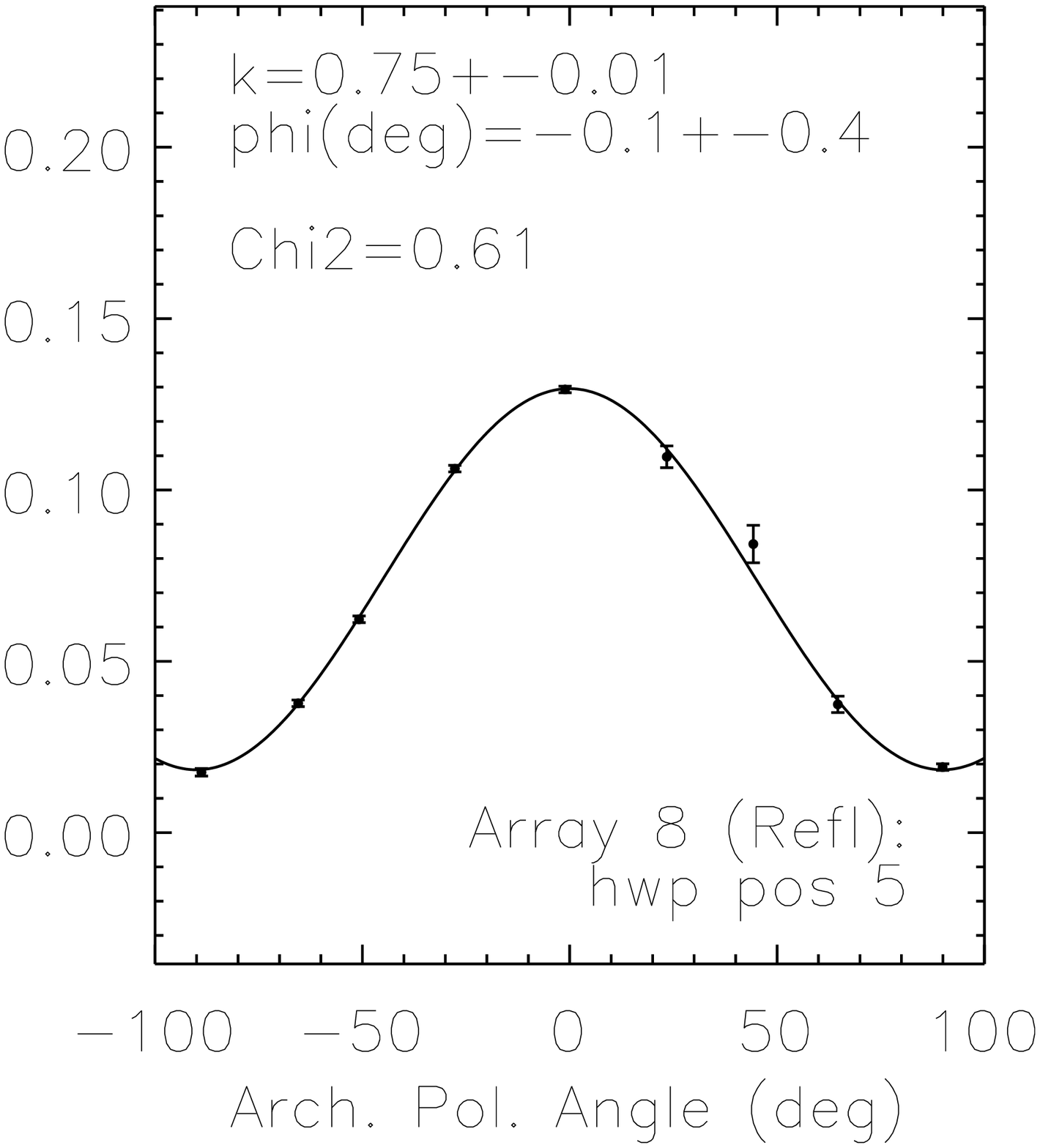}
\includegraphics[width=4.2cm,trim=4cm 0 0 0]{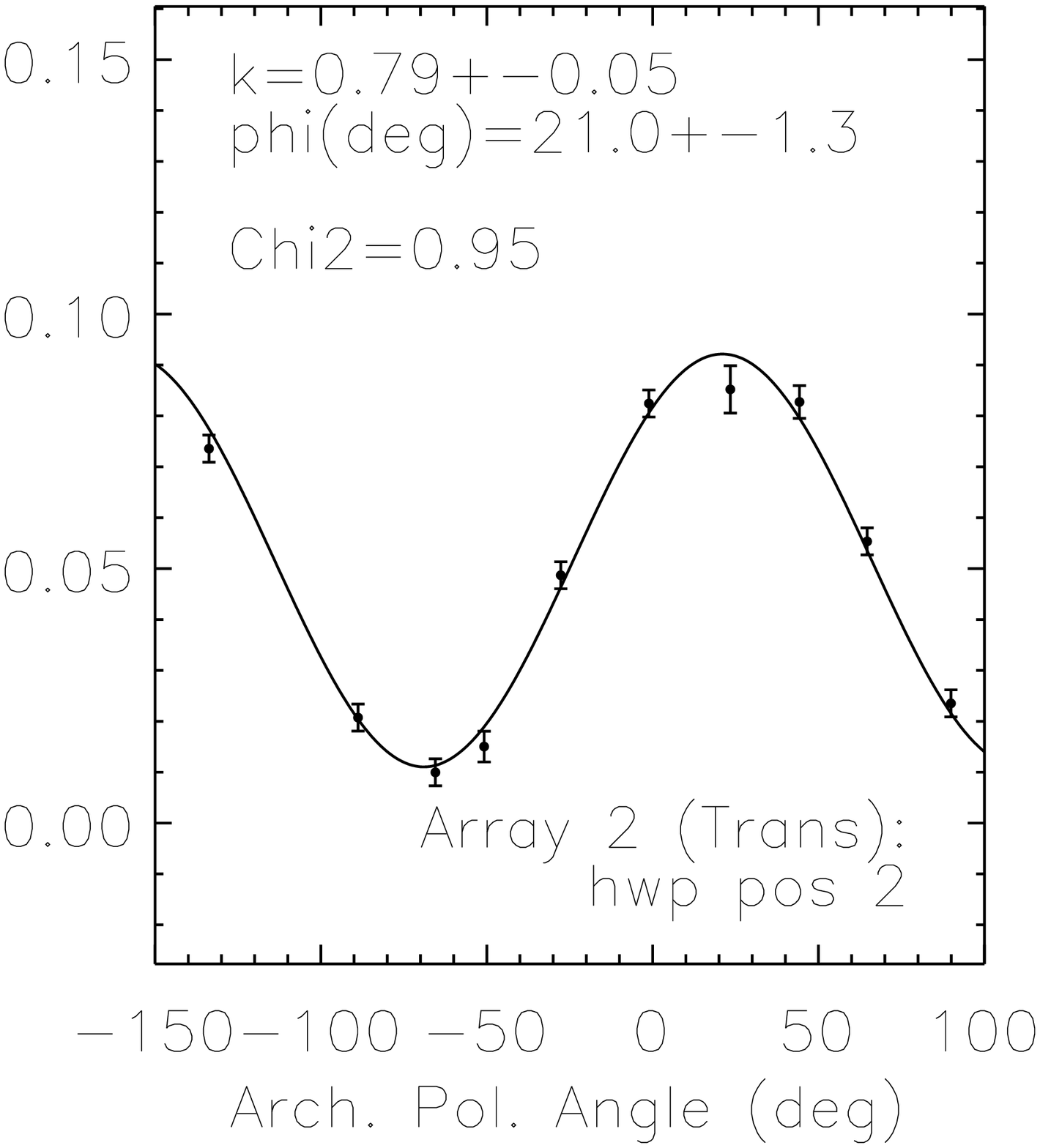}
\includegraphics[width=4.2cm,trim=4cm 0 0 0]{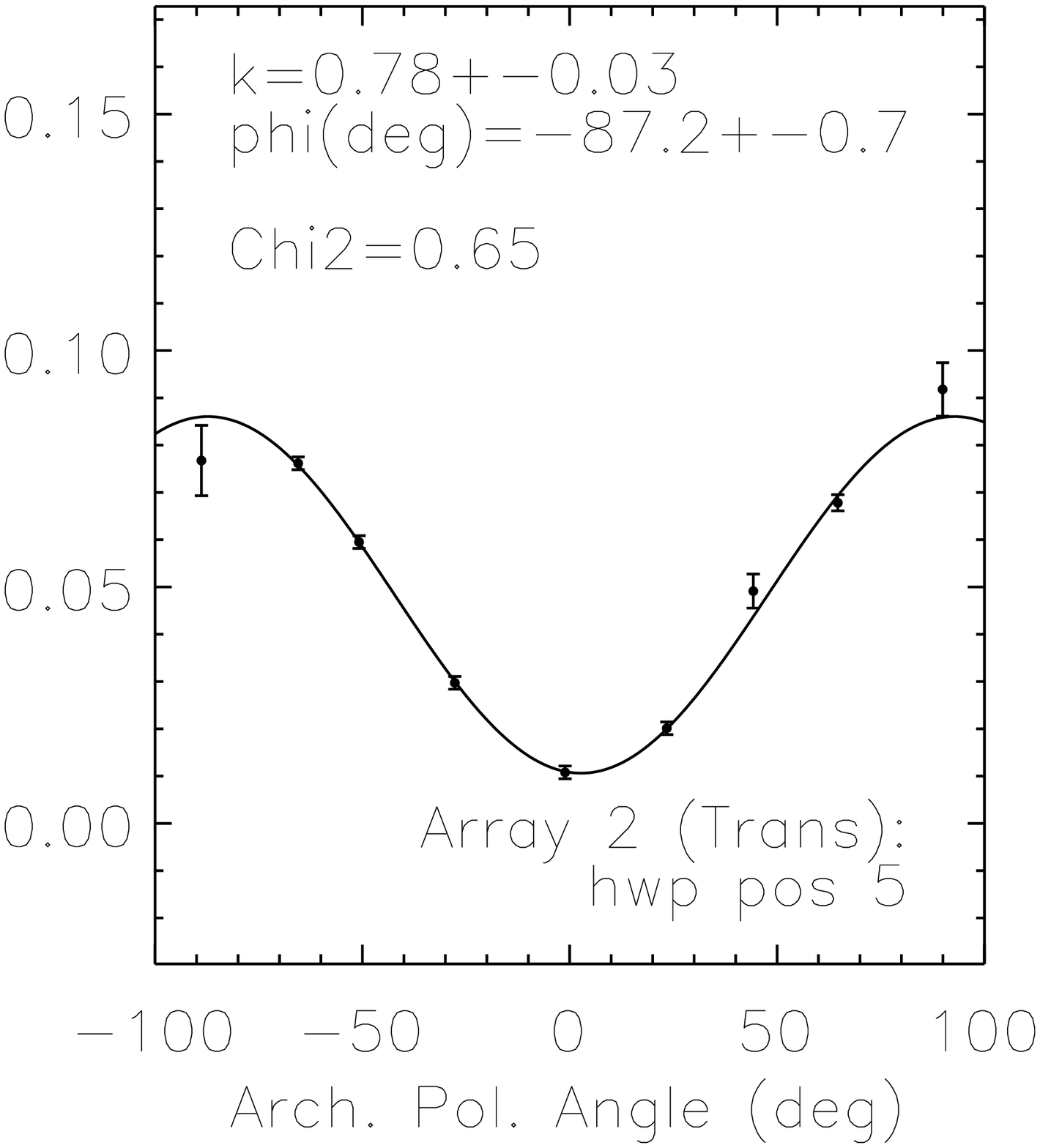}
\caption{Peak flux of the PSF fitted on arrays 2 and 8 for positions 2
($-40.25\degr$) and 5 ($-2.75\degr$) of the HWP, as a function of the nine
angular positions of \archpol. Error bars have been derived a posteriori.}
\label{fig:arch_pol}
\end{center}
\end{figure}

These observations for two positions of the HWP and nine positions of
\archpol\ are presented on Fig.~\ref{fig:arch_pol}. The best fit shown
in the figure are consistent with a polarization fraction of \archpol \ 
between $k=0.75\%$ and $k=0.79\%$.  Measures of \archpol's
cross-polarization $k$ are consistent for each HWP position on a given
array but marginally consistent between the two arrays when the HWP is
in position 5 and uncertainties on $k$ are best.

\subsubsection{HALF WAVE PLATE CHARACTERIZATION}
\label{sec:HalfWavePlateCharacterization}

\begin{figure}
\begin{center}
\includegraphics[clip, angle=0, scale = 0.25]{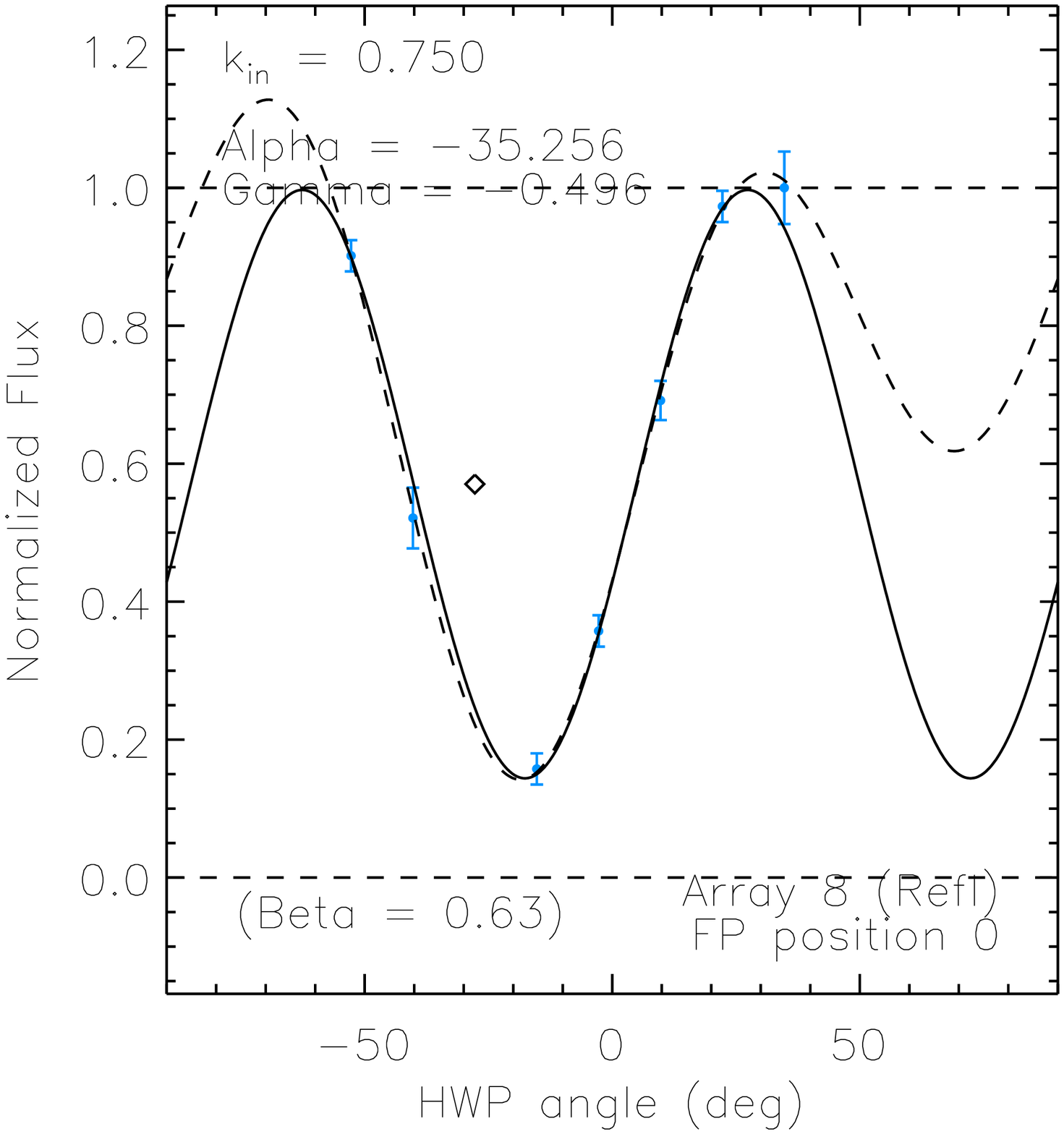}
\includegraphics[clip, angle=0, scale = 0.25]{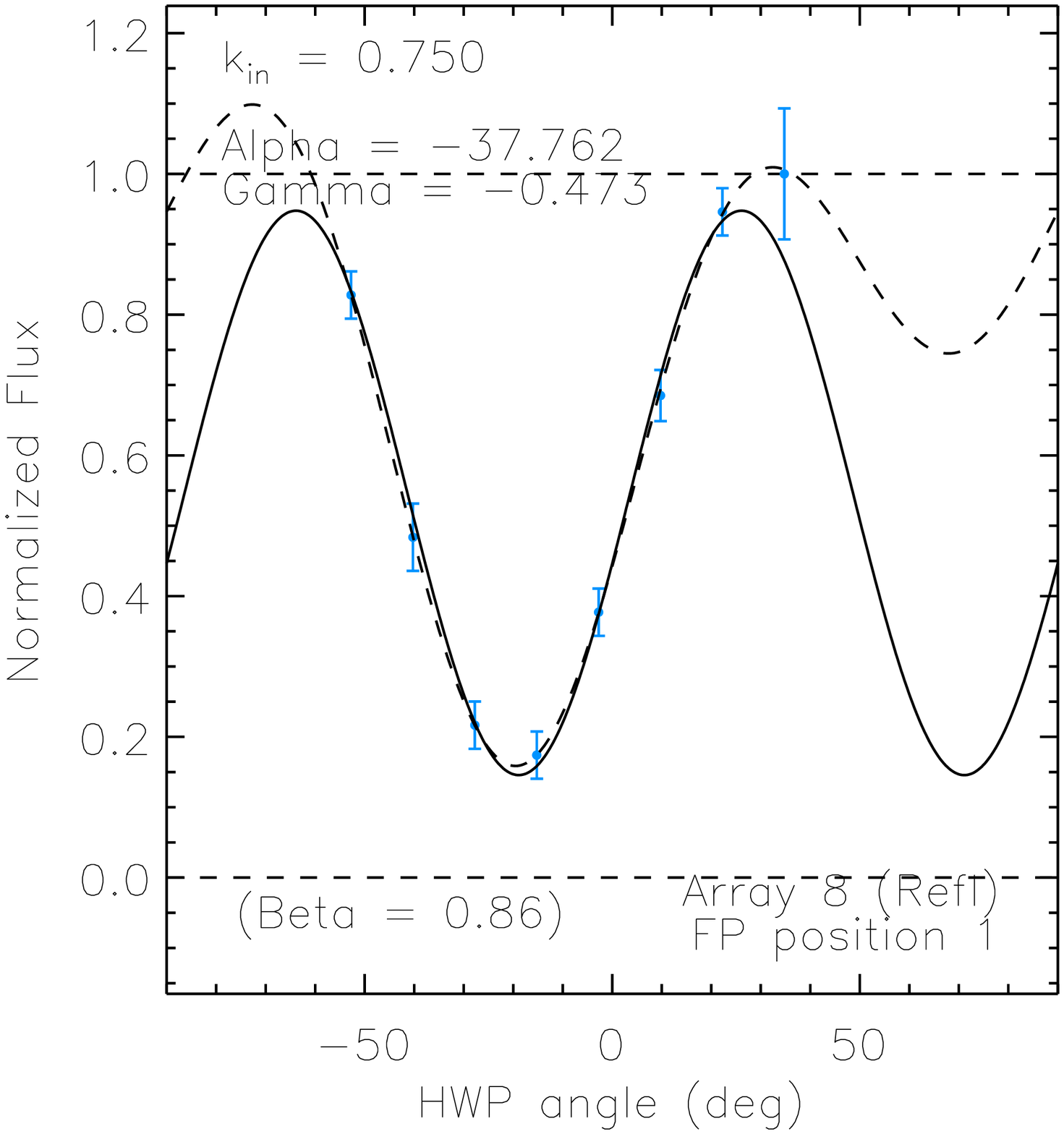}
\includegraphics[clip, angle=0, scale = 0.25]{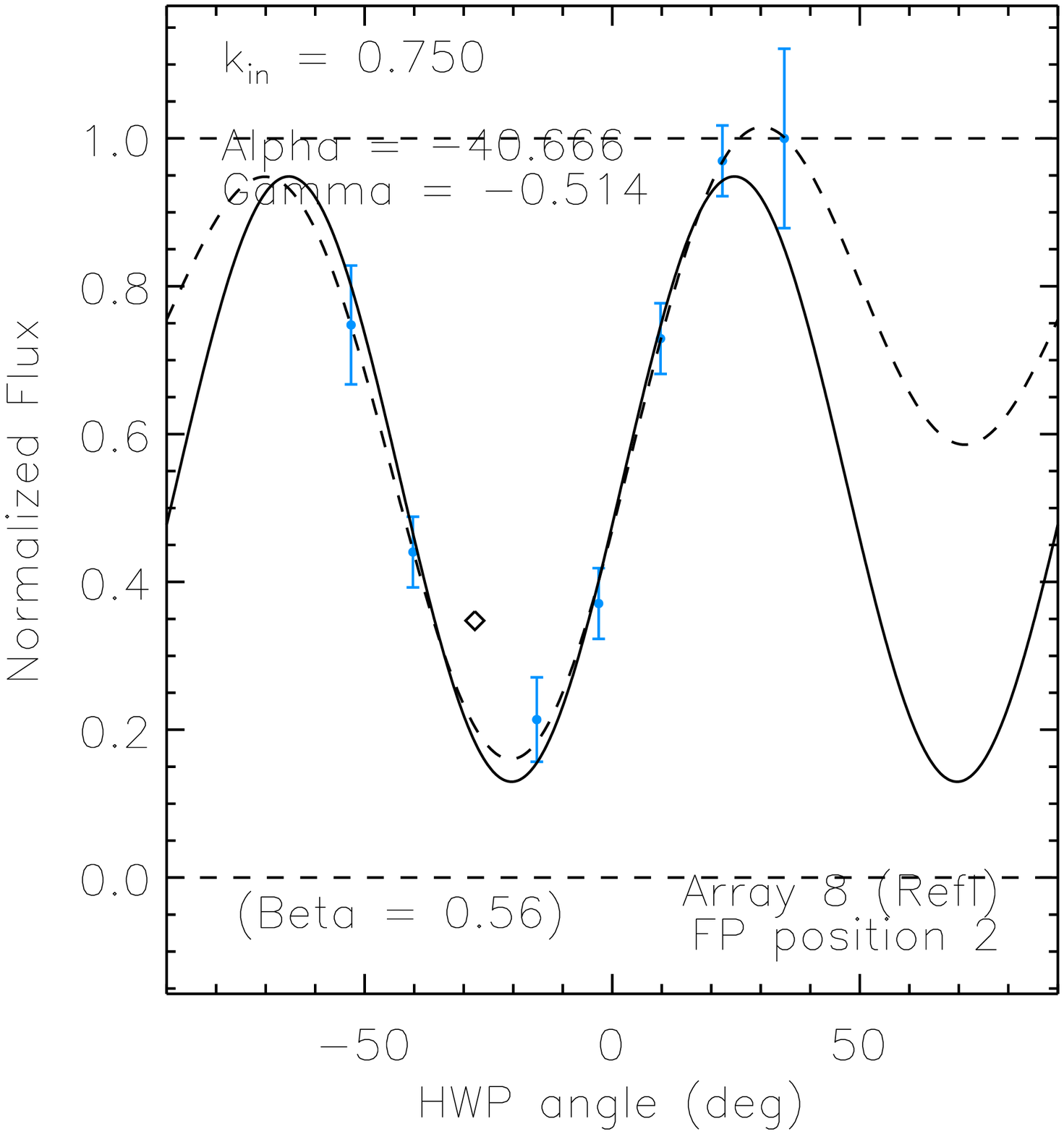}
\includegraphics[clip, angle=0, scale = 0.25]{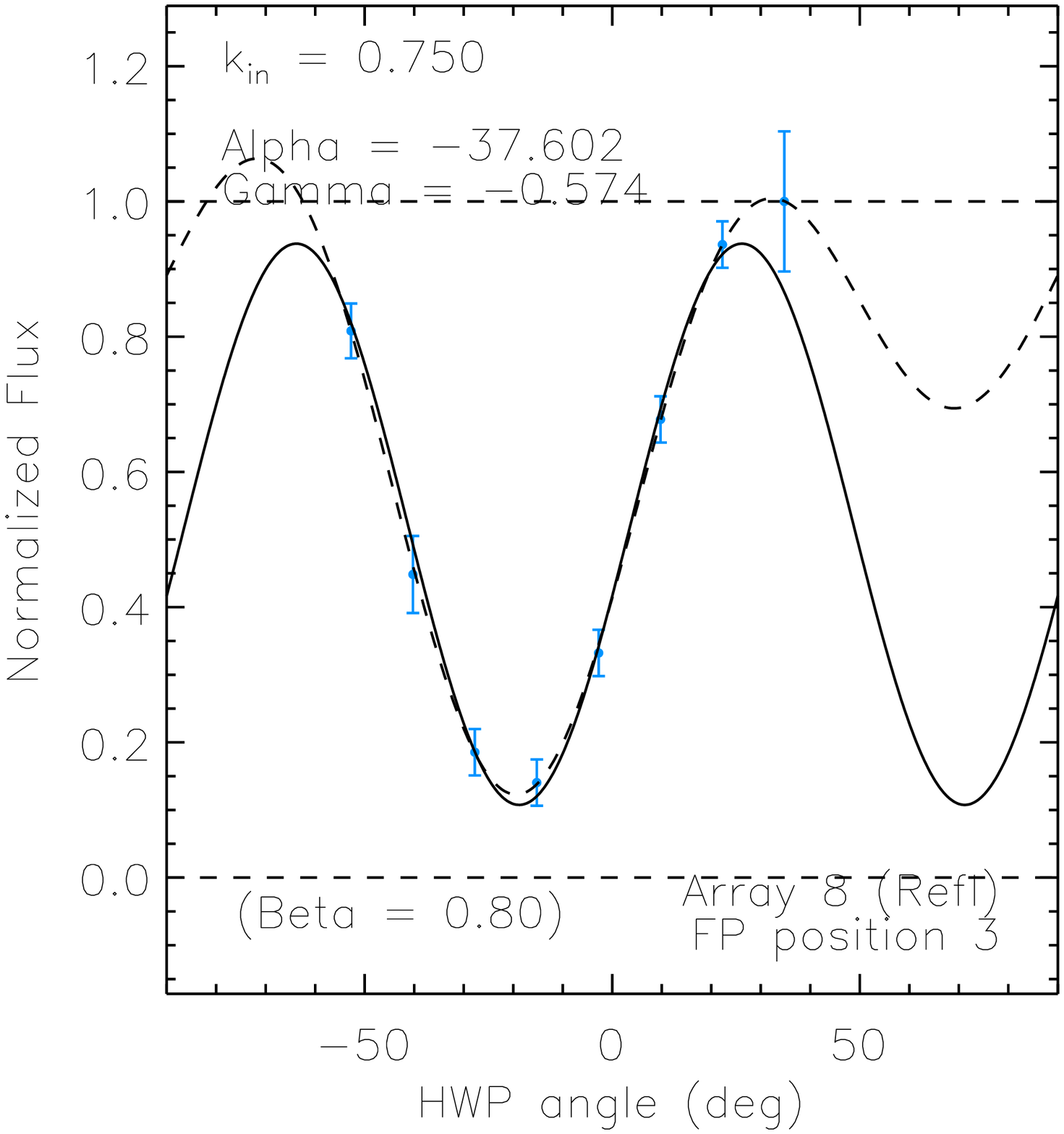}
\includegraphics[clip, angle=0, scale = 0.25]{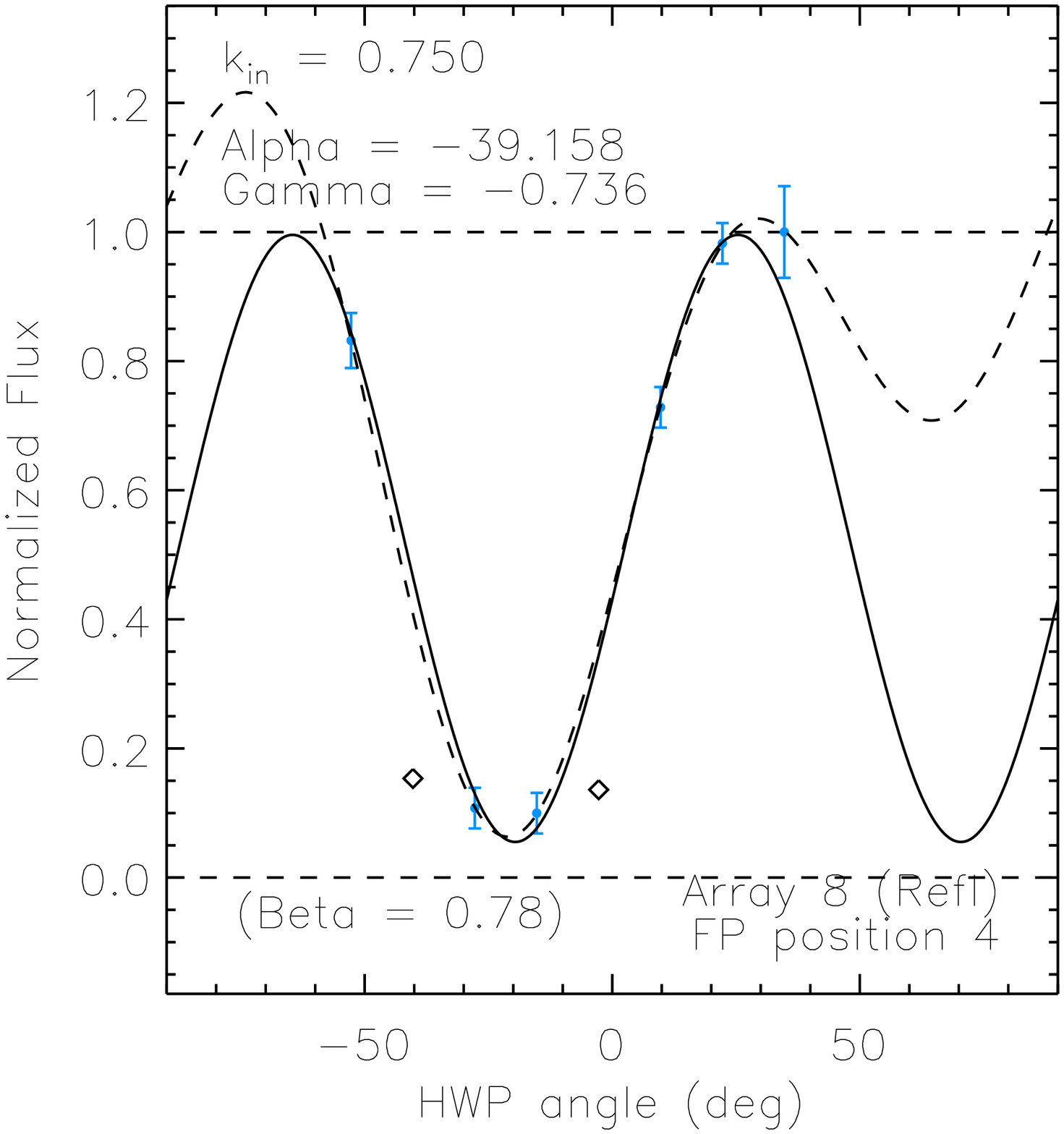}
\includegraphics[clip, angle=0, scale = 0.25]{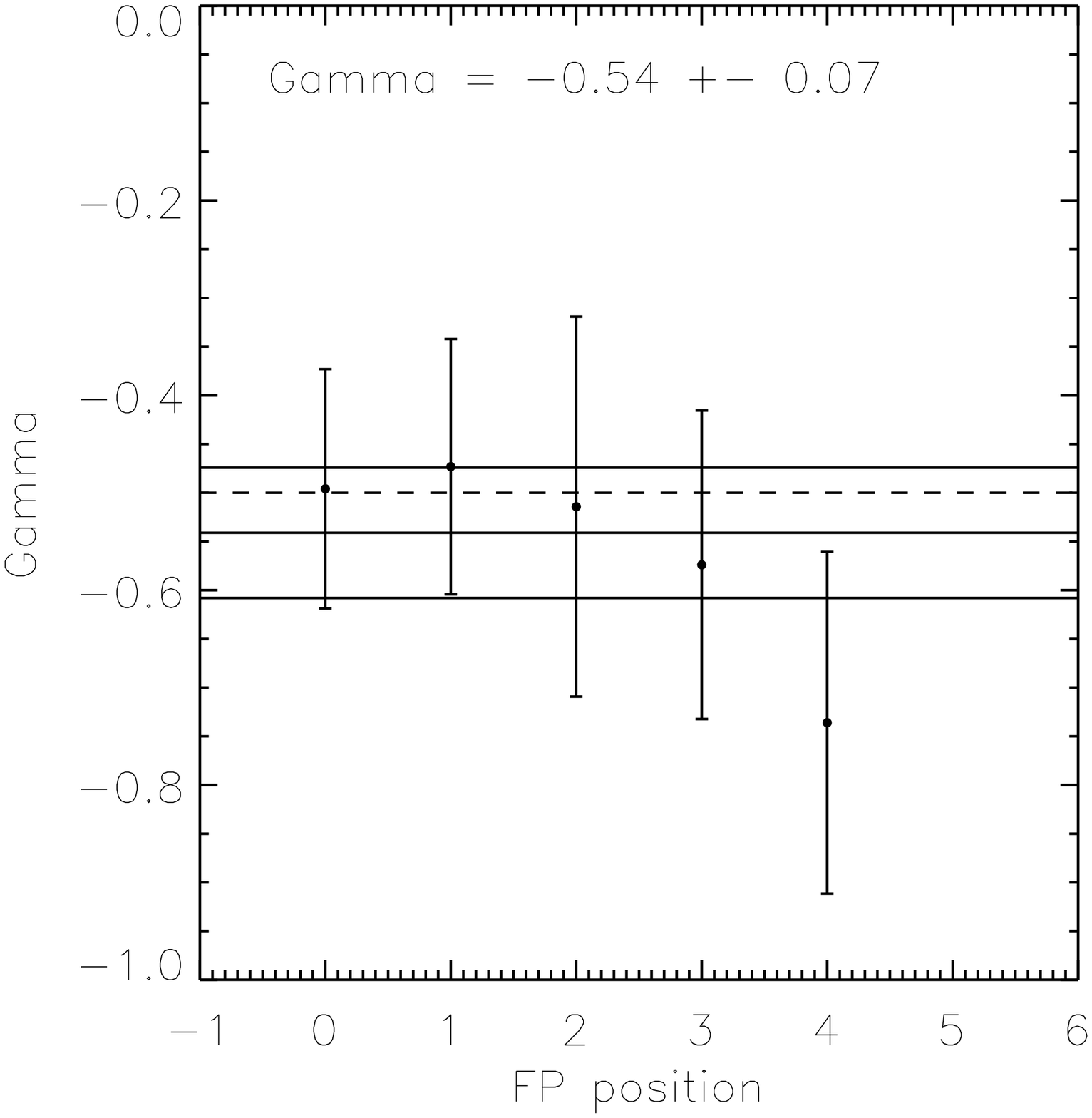}
\caption{Peak flux of the PSF fitted on array 8 for 5 positions on the array as a
function of the HWP angle. Error bars have been derived a posteriori. Data
points marked as black empty diamonds have been discarded for the fit. On the
last plot, the dash line indicates the theoretical perfect value $\gamma=-0.5$, the
solid lines are the average value for $\gamma$ and its one $\sigma$ uncertainty.
\label{fig:hwp_array8}}
\end{center}
\end{figure}

\begin{figure}
\begin{center}
\includegraphics[clip, angle=0, scale = 0.25]{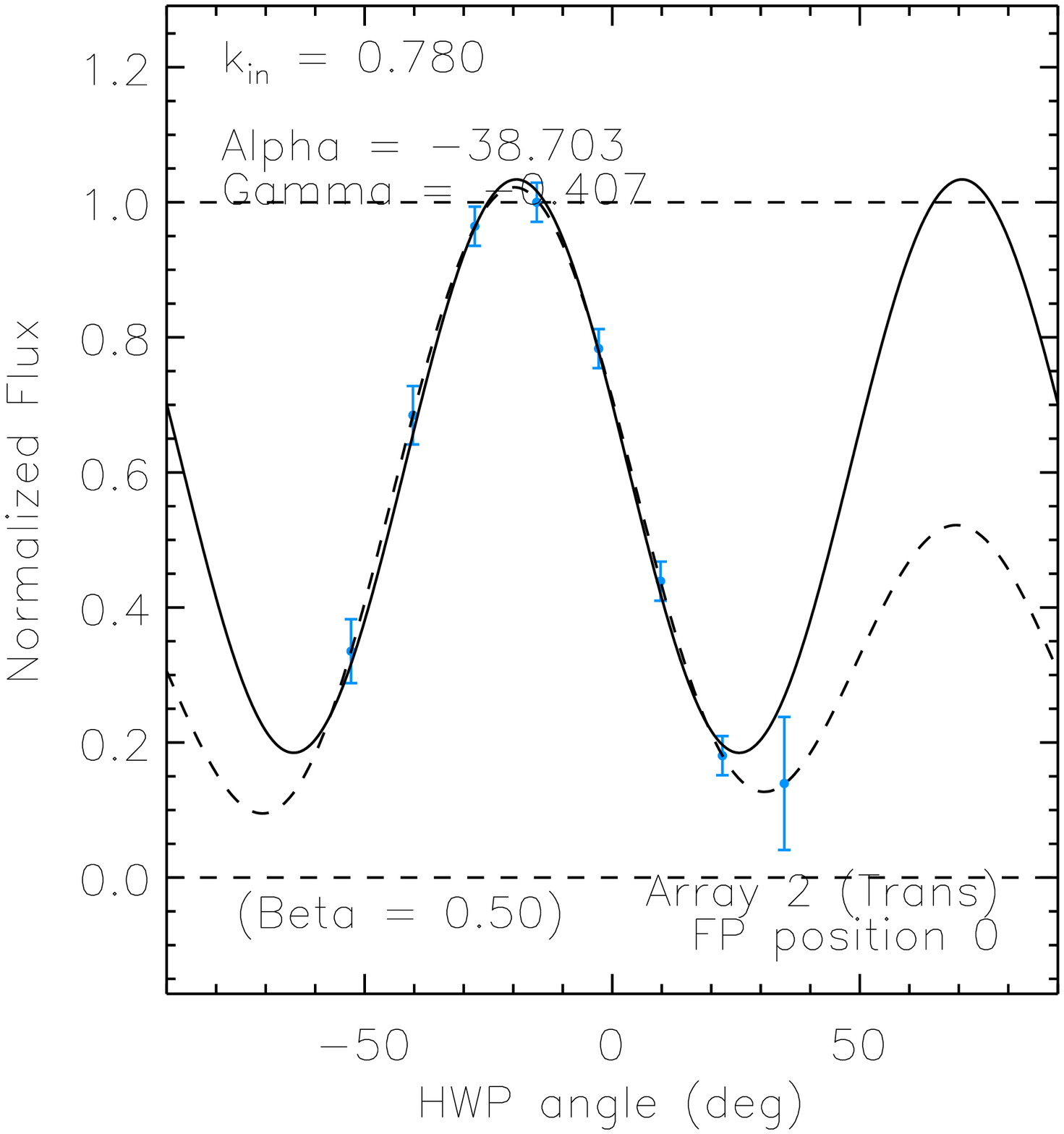}
\includegraphics[clip, angle=0, scale = 0.25]{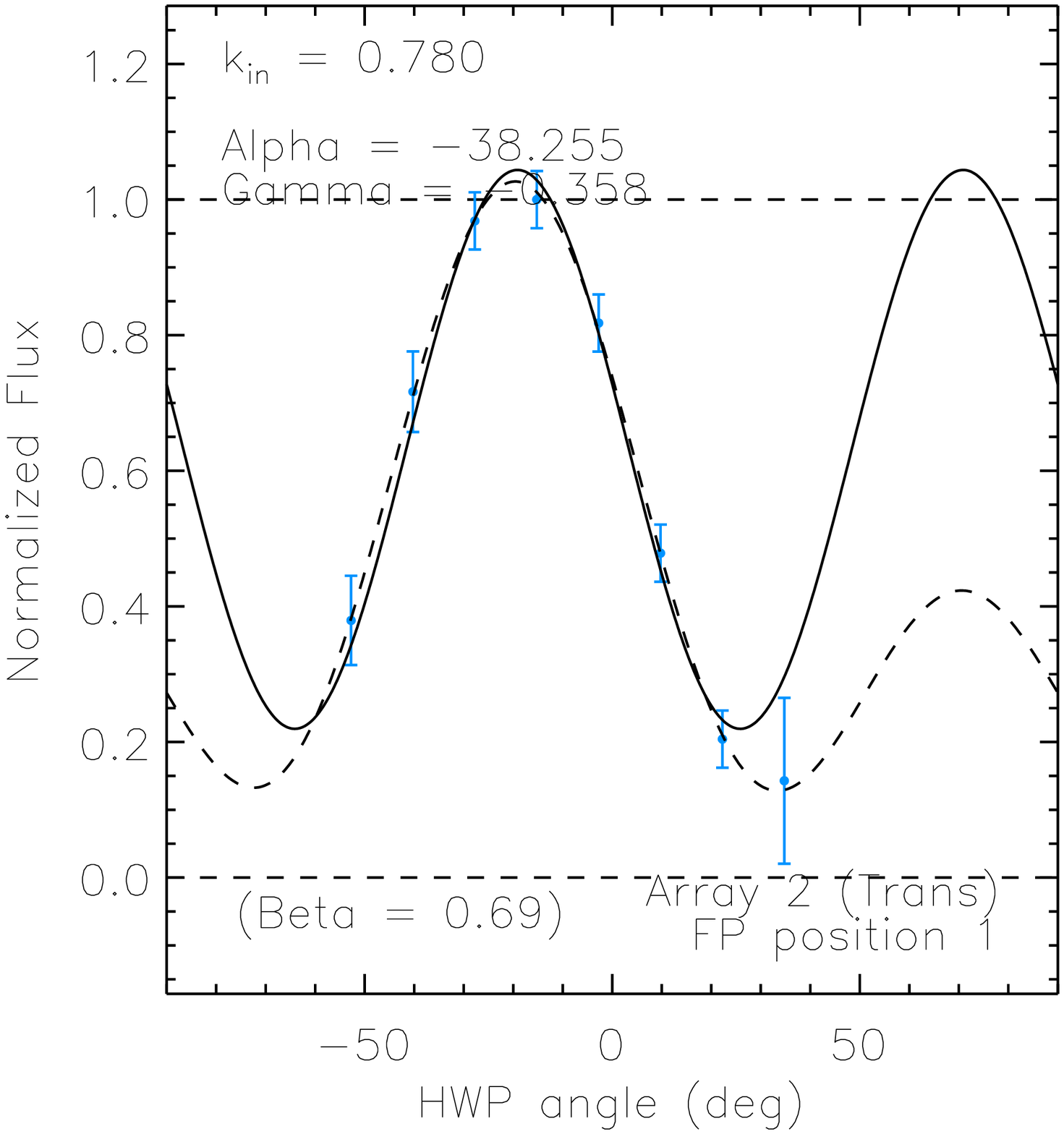}
\includegraphics[clip, angle=0, scale = 0.25]{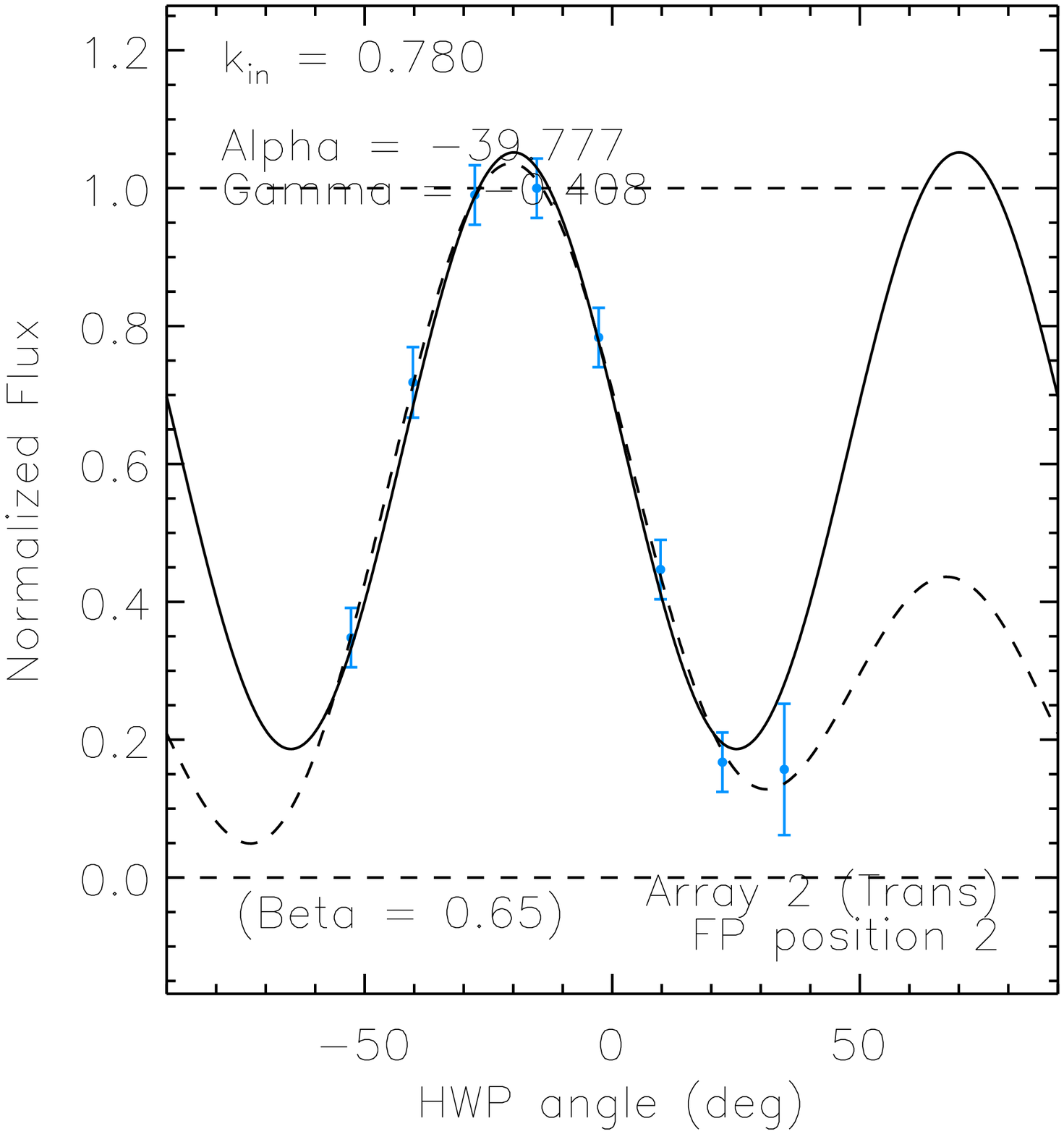}
\includegraphics[clip, angle=0, scale = 0.25]{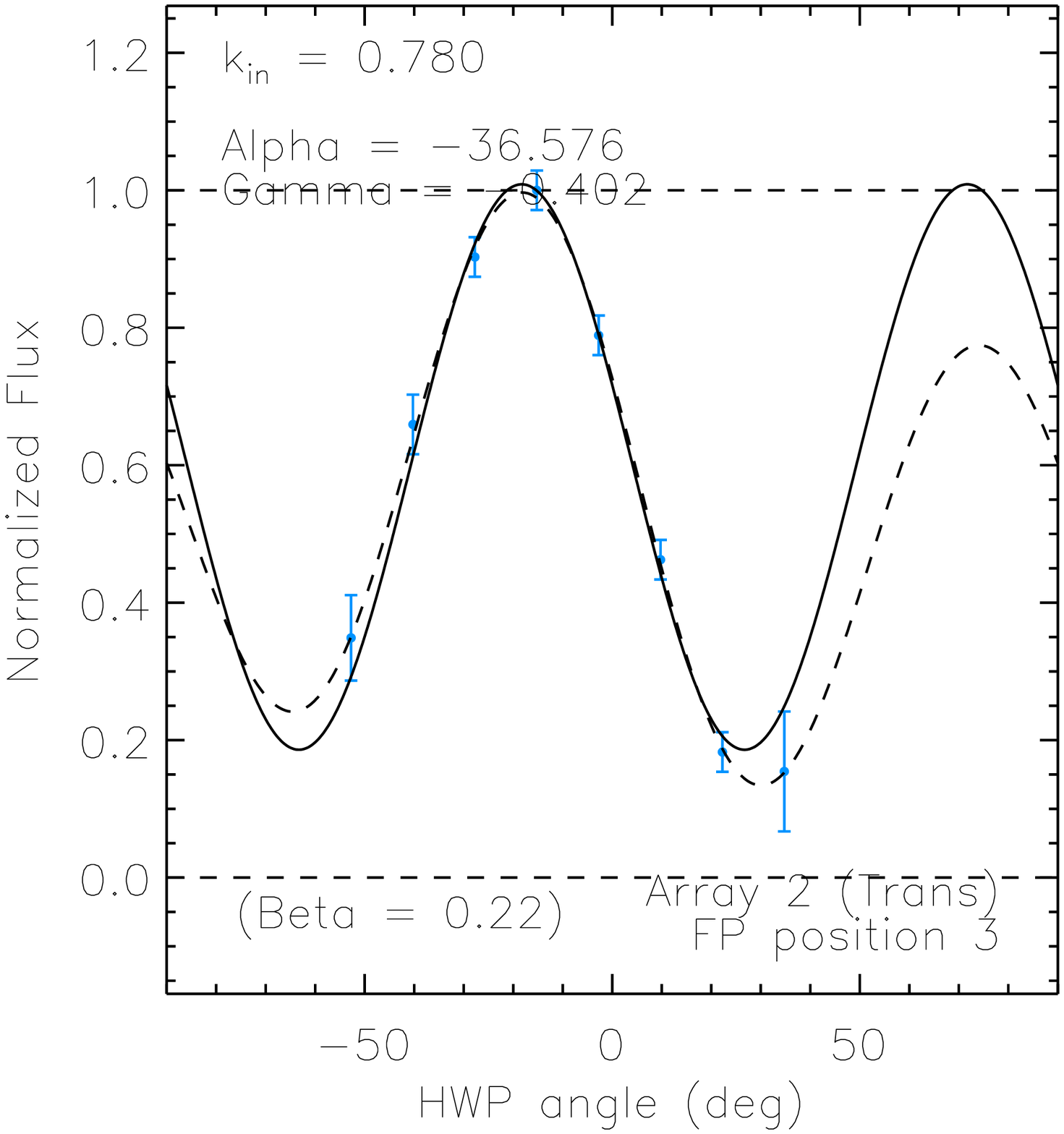}
\includegraphics[clip, angle=0, scale = 0.25]{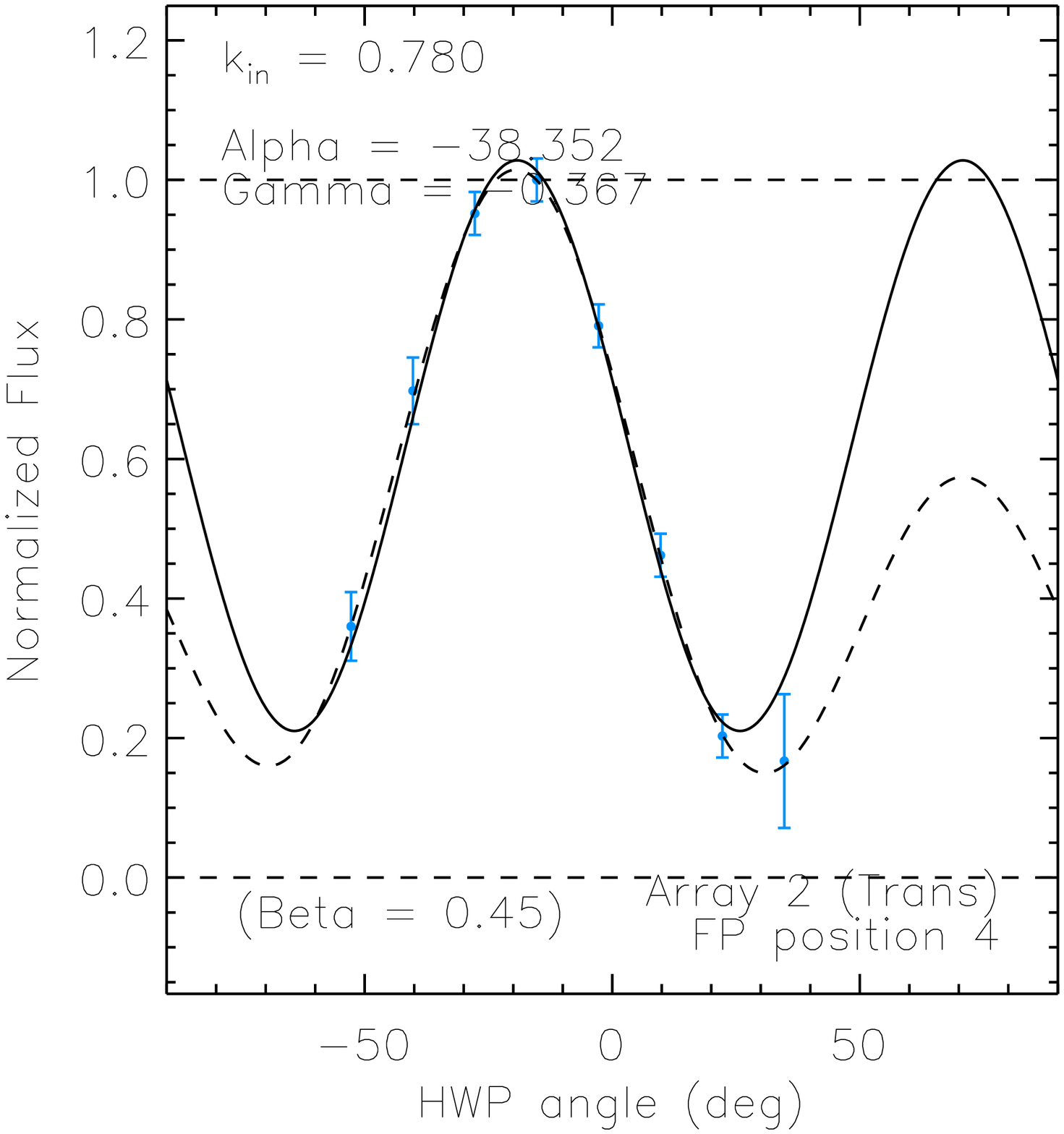}
\includegraphics[clip, angle=0, scale = 0.25]{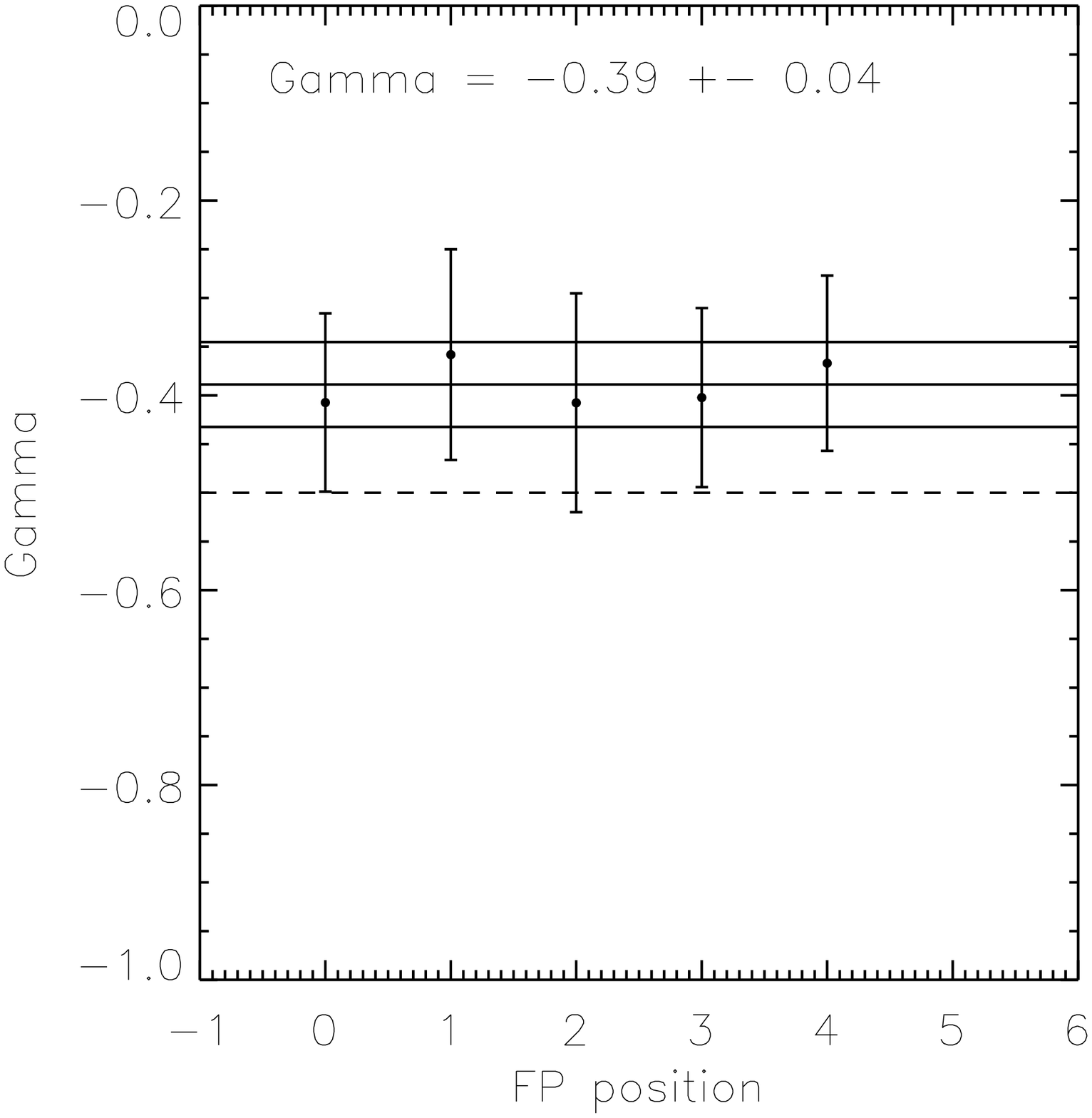}
\caption{Peak flux of the PSF fitted on array 2 for 5 positions on the array as a
function of the HWP angle. Error bars have been derived a posteriori. Data
points marked as black empty diamonds have been discarded for the fit. On the
last plot, the dash line indicates the theoretical perfect value $\gamma=-0.5$, the
solid lines are the average value for $\gamma$ and its one $\sigma$ uncertainty.
\label{fig:hwp_array2}}
\end{center}
\end{figure}

We take the values of $k$ derived for the transmission and the reflection arrays
in section~\ref{se:arch_pol} and now fit the peak of the PSF as a function of
the HWP angle for a fixed orientation of \archpol (Figs.\ref{fig:hwp_array8}
and \ref{fig:hwp_array2}). If the HWP is good, $\beta$ should be close to zero, there
should be no term in $2\omega$ in the fit should be good with terms in $4\omega$
only. This can be seen on the solid line fits. Allowing for terms in $2\omega$ (dash
line) leads to inconsistent determinations of $\beta$ with no significant
improvements of the goodness of fit (although error bars are still determined a
posteriori here). This, together with the finding that $\gamma = -0.5 \pm 0.04$
using the transmission array and $\gamma = -0.49 \pm 0.02$ using the reflection
array indicates that the HWP is indeed performing as well as expected.

We fit the observed source fluxes independently for the Reflection and
Transmission arrays, using
\begin{equation}
\label{eq:measure}
m_T = a_0 + a_1\cos4\omega + a_2\sin4\omega,
\end{equation}
where $a_0$, $a_1$ and $a_2$ are free parameters, related to physical
parameters through
\begin{eqnarray}
\alpha   & = & \frac{1}{2}\arctan\frac{a_2}{a_1} \\
\rho_{T,R}     & = & \frac{\max(m_{T,R})-\min(m_{T,R})}{\max(m_{T,R})+\min(m_{T,R})} = \frac{a_0}{\sqrt{a_1^2+a_2^2}}\\
\gamma_T & = & \frac{2-k\rho_T+k\cos2\alpha}{-2k\rho_T-2k\cos2\alpha}\\
\gamma_R & = & \frac{2-k\rho_R-k\cos2\alpha}{-2k\rho_R+2k\cos2\alpha}
\end{eqnarray}
The best fit are shown in Fig.\,\ref{fig:hwp_array8} and
Fig.\,\ref{fig:hwp_array2}.

No evidence has been found for a significant differential transmission
between the fast and slow axis of the HWP ($\beta=0$) and no evidence has been
found either for a defect related to the phase shift ($\gamma = -0.5 \pm
0.04$ or $-0.49 \pm 0.02$).

\subsubsection{POLARIZATION ROTATION}

Optical calculations show that the direction of an incident linear
polarization can be slightly rotated through the optics, as it
propagates from the primary mirror to the detector. This rotation
angle varies across the focal-plane, and, according to calculations,
can reach a few degrees away from the optical axis. This rotation must
be measured precisely and taken into account during astronomical data
processing.  To characterize this rotation, the whole focal plane has
been scanned with the polarizer set at different polarization
directions, in order to measure the polarization distortion. For each
position, the direction of the polarizer has been measured as
described above.

\begin{figure}[htb]
\centering
\includegraphics[width=8cm]{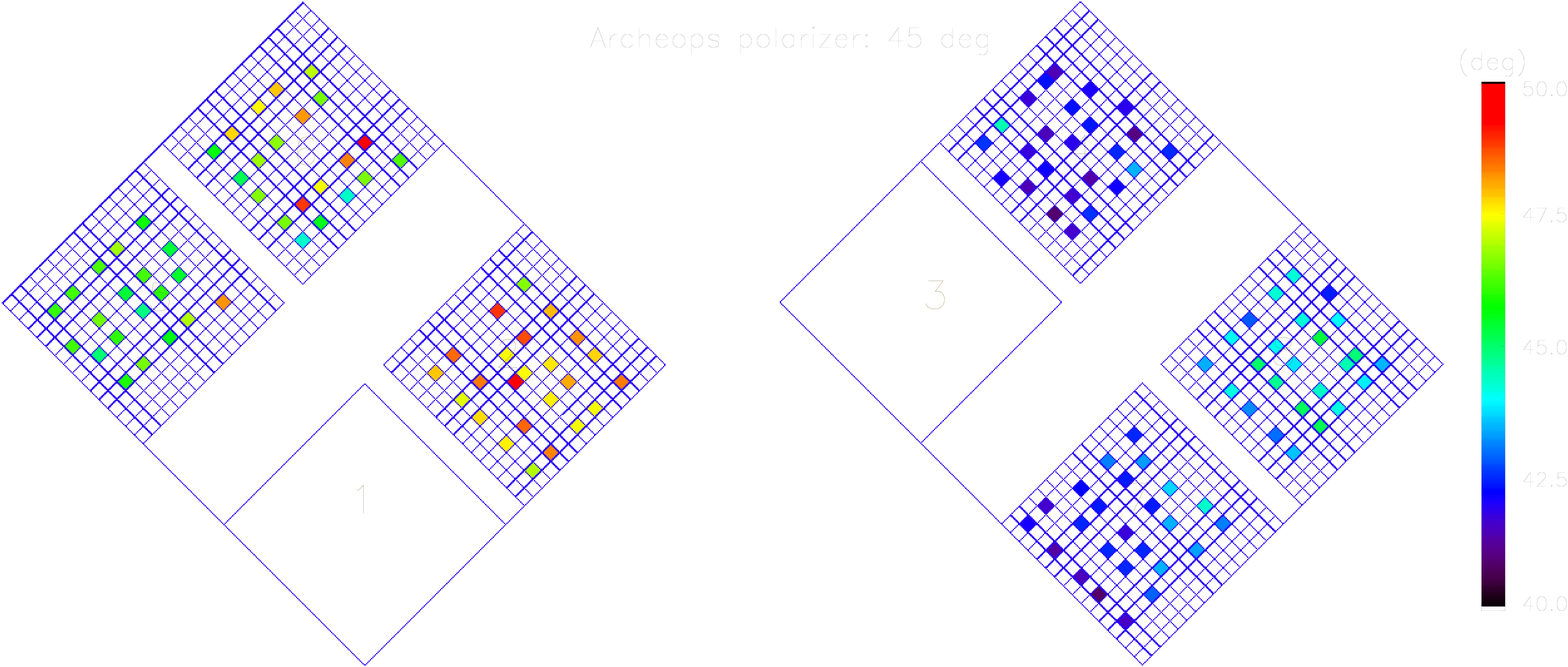}
\includegraphics[width=8cm]{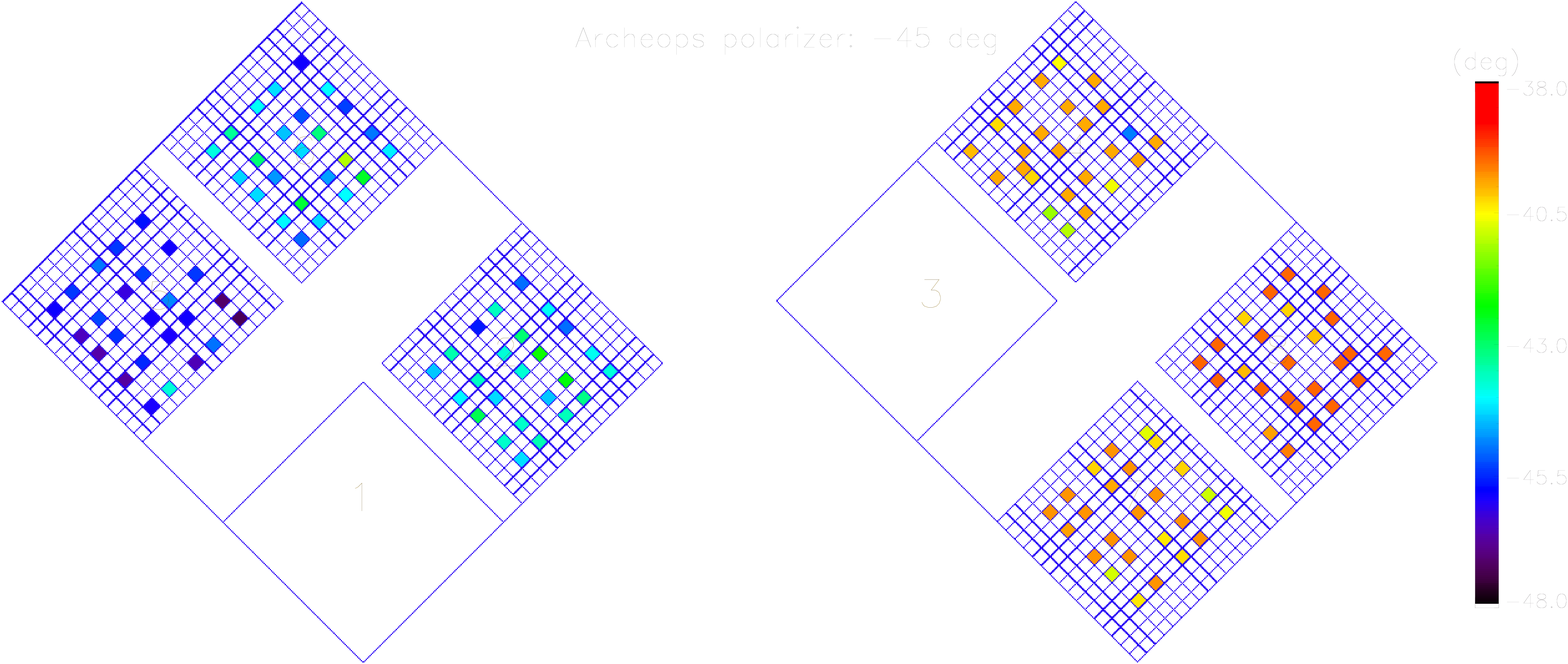}
\caption{
Map of the recovered polarization angles for
point sources placed at various locations  in the \Pilot focal
plane. The input linear polarization direction was $45.19 \pm 0.30\degr$ (left) and 
 $-44.81 \pm 0.30\degr$ (right).
\label{fig:pol_rot}}
\end{figure}

\begin{figure}[htb]
\centering
\includegraphics[width=6cm]{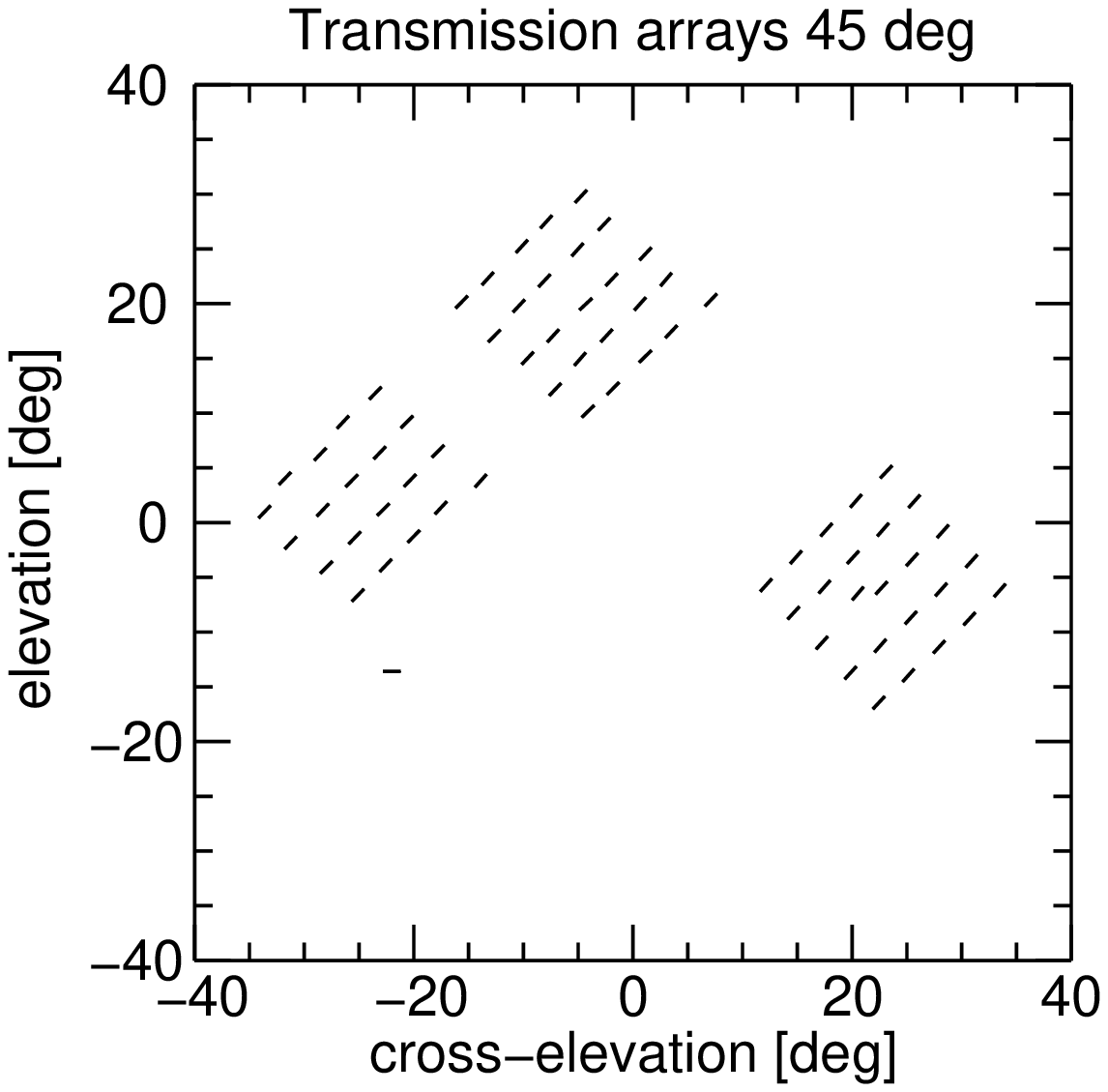}
\includegraphics[width=6cm]{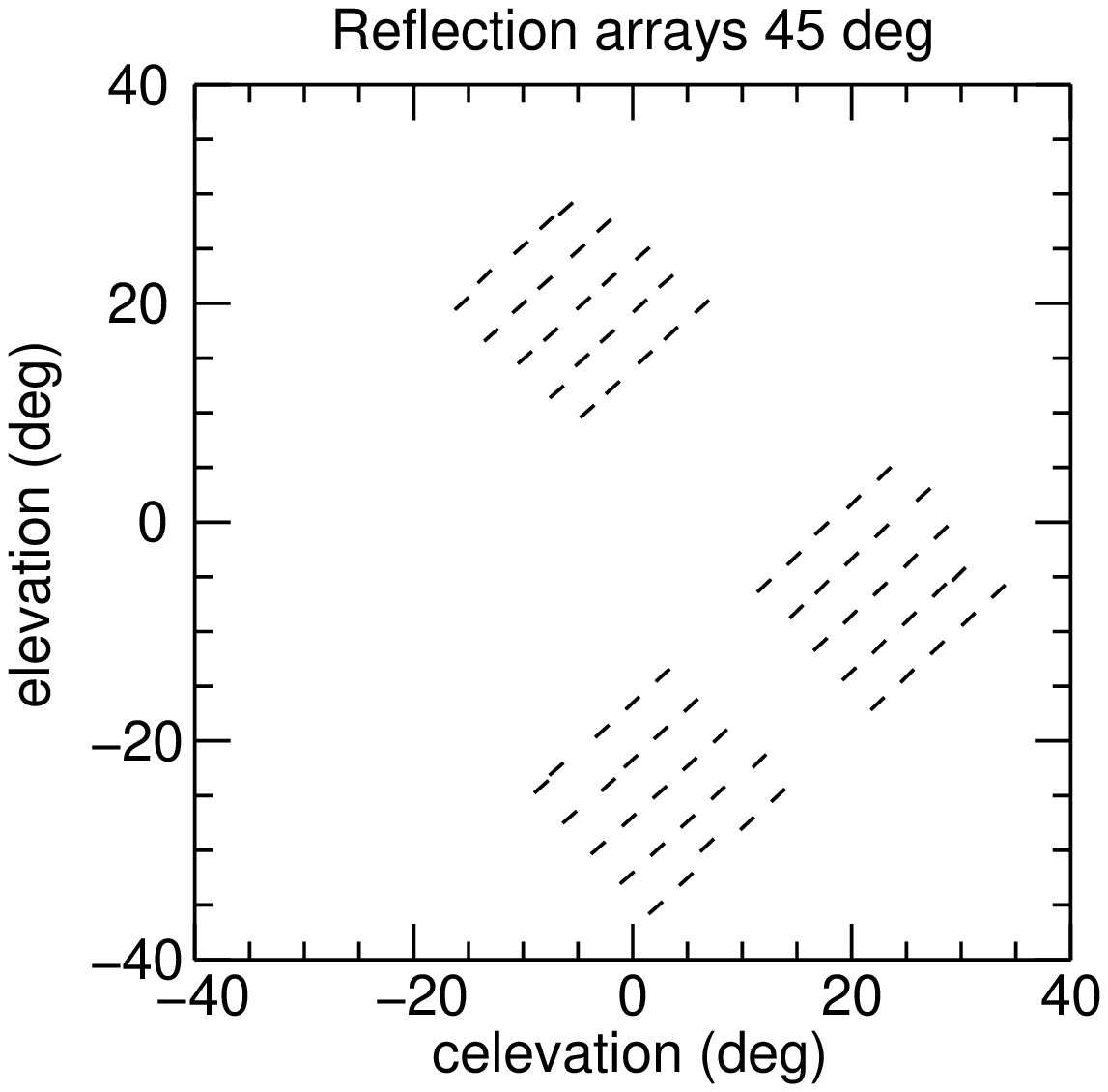}
\includegraphics[width=6cm]{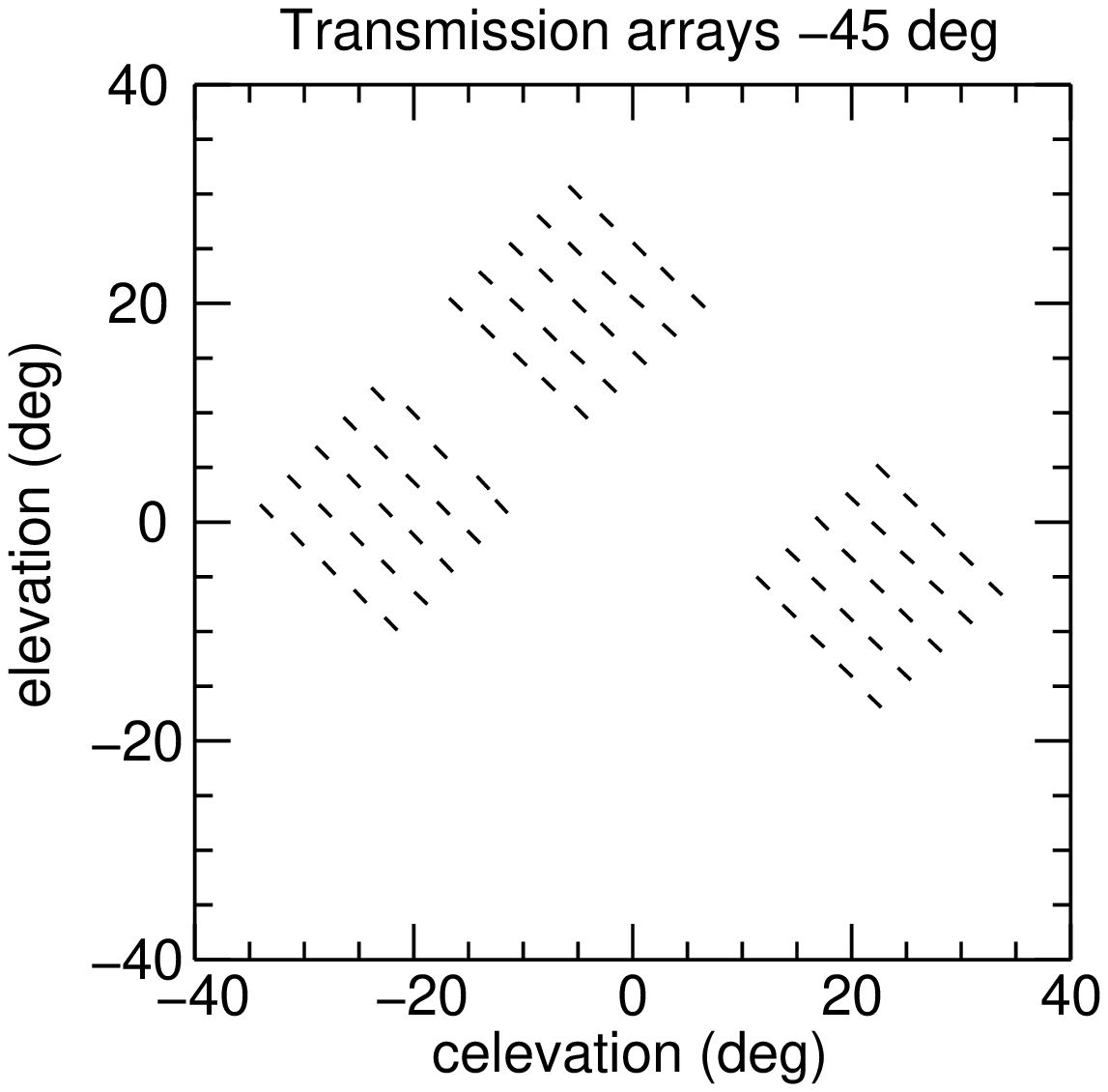}
\includegraphics[width=6cm]{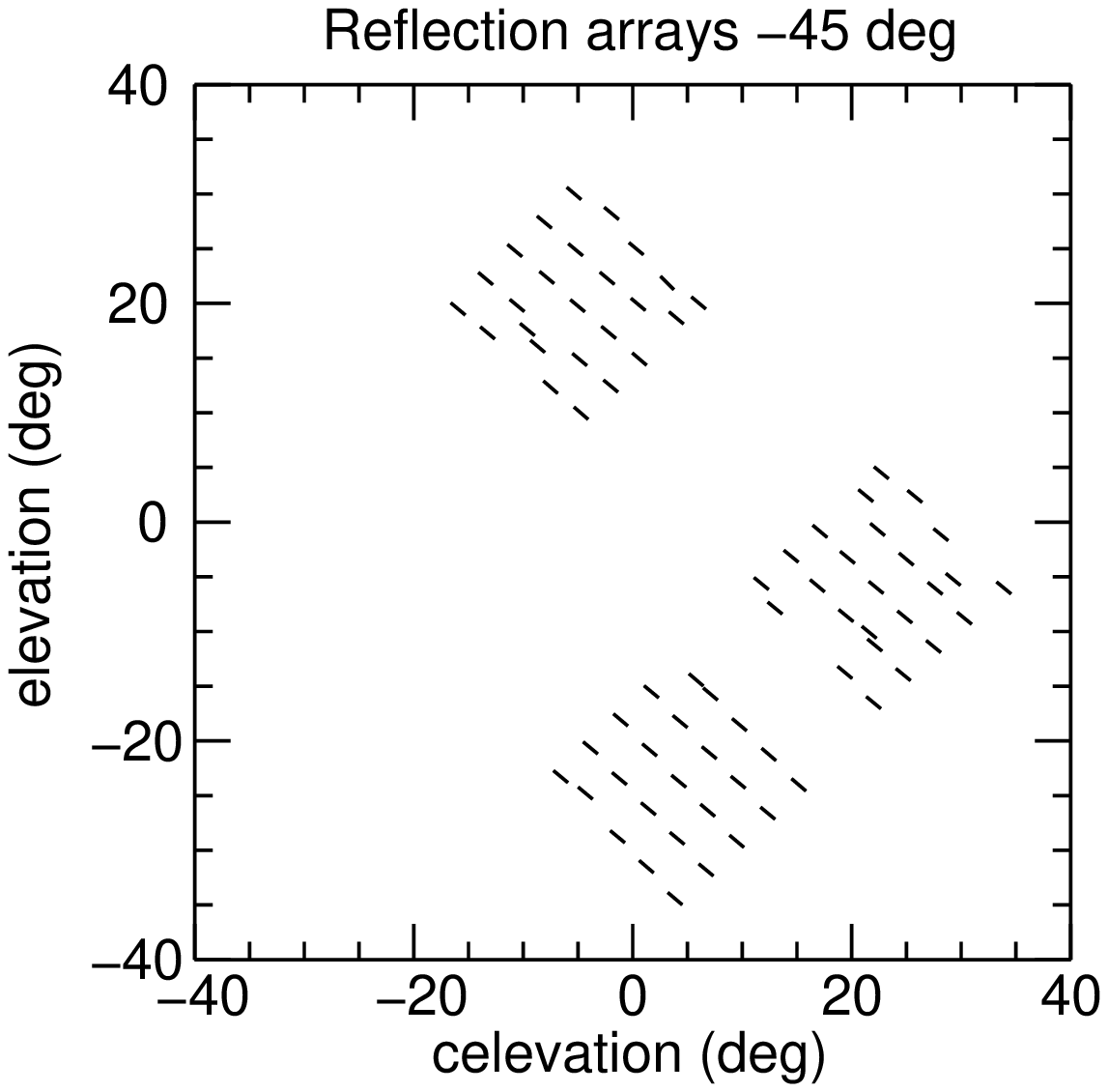}
\caption{
Map of the recovered polarization angles for
point sources placed at various locations  in the \Pilot focal
plane. The input linear polarization direction was $45.19 \pm 0.30\degr$ (left) and 
 $-44.81 \pm 0.30\degr$ (right).
\label{fig:pol_rot_vec}}
\end{figure}

We used the polarizer at angles $45.19 \pm 0.30\degr$ deg and $-44.81
\pm 0.30\degr$ and moved the polarized collimator point source over 25
pixels of each array. At each position, we rotated the HWP over its
full angle range (8 positions). The data is averaged for each PSF
position and the PSF flux as a function of HWP position is fitted
using Eq.\,\ref{eq:measure} in order to derive polarization
parameters.  Figure\,\ref{fig:pol_rot} shows the distribution of
polarization angles over the focal plane. It can be seen that the
polarization direction is recovered with the right angle. The actual
average recovered values are $45.39\degr$ and $-42.6\degr$. We evidence
a systematic rotation across the focal plane with an amplitude of about
$3\degr$ (see Fig.\,\ref{fig:pol_rot_vec}).


\section{CONCLUSIONS}

We have performed a series of ground test of the \Pilot experiment,
using a collimated source and a large diameter polarizer. We used this
system to control the optical performances of the instrument.
We measured the width of the PSF of the instrument as a function of
defocusing of the primary mirror and found the optimum defocus. The
analysis of the PSF shape shows that the optical system conforms to
expectations. We also used the tests to reconstruct
accurately measure the geometry of the focal plane. Results in
polarization indicate that the large scale polarizer used has a 73\,\%
polarization efficiency at $240\mic$. We tested the performances of
the Half-Wave plate internal to the \Pilot photometer and found no
evidence for deviation from perfection. We measured the focal plane distribution
of polarization directions for a given direction of the point source
polarization and evidenced small rotation by the optical system.
We used background measurements at various temperatures to separate
optical vignetting and electronics or detector offsets.
These tests demonstrated that the optics of the \Pilot system meets
requirements. The data obtained during those tests will be critical to
analyze the first scientific data, that will be obtained during the
first stratospheric balloon flight of the instrument, in 2015.

\bibliography{SPIE2014}   

\begin{thebibliography}{10}

\bibitem{page2007}
{Page}, L., {Hinshaw}, G., {Komatsu}, E., {Nolta}, M.~R., {Spergel}, D.~N.,
  {Bennett}, C.~L., {Barnes}, C., {Bean}, R., {Dor{\'e}}, O., {Dunkley}, J.,
  {Halpern}, M., {Hill}, R.~S., {Jarosik}, N., {Kogut}, A., {Limon}, M.,
  {Meyer}, S.~S., {Odegard}, N., {Peiris}, H.~V., {Tucker}, G.~S., {Verde}, L.,
  {Weiland}, J.~L., {Wollack}, E., and {Wright}, E.~L., ``{Three-Year Wilkinson
  Microwave Anisotropy Probe (WMAP) Observations: Polarization Analysis},''
  {\em ApJS}~{\bf 170},  335--376 (June 2007).

\bibitem{fosalba2002}
{Fosalba}, P., {Dor{\'e}}, O., and {Bouchet}, F.~R., ``{Elliptical beams in CMB
  temperature and polarization anisotropy experiments: An analytic approach},''
  {\em PhRvD}~{\bf 65},  063003 (Mar. 2002).

\bibitem{Hildebrand1999}
{Hildebrand}, R.~H., {Dotson}, J.~L., {Dowell}, C.~D., {Schleuning}, D.~A., and
  {Vaillancourt}, J.~E., ``{The Far-Infrared Polarization Spectrum: First
  Results and Analysis},'' {\em ApJ}~{\bf 516},  834--842 (May 1999).

\bibitem{Benoit2004a}
{Beno{\^i}t}, A. and {Archeops Collaboration}, ``{ARCHEOPS: a balloon
  experiment for measuring the cosmic microwave background anisotropies},''
  {\em Advances in Space Research}~{\bf 33},  1790--1792 (Jan. 2004).

\bibitem{Benoit2004}
{Beno{\^i}t}, A., {Ade}, P., {Amblard}, A., {Ansari}, R., {Aubourg}, {\'E}.,
  and {et al.}, ``{First detection of polarization of the submillimetre diffuse
  galactic dust emission by Archeops},'' {\em A\&A}~{\bf 424},  571--582 (Sept.
  2004).

\bibitem{planck2014-XXI}
{\sorthelp{Planck Collaboration Int U}}{Planck Collaboration Int. XXI},
  ``{\Planck\ intermediate results. XXI. Comparison of polarized thermal
  emission from Galactic dust at 353\GHz\ with optical interstellar
  polarization},'' {\em A\&A, submitted, [arXiv:astro-ph/1405.0873]}  (2014).

\bibitem{planck2014-XIX}
{\sorthelp{Planck Collaboration Int S}}{Planck Collaboration Int. XIX},
  ``{\Planck\ intermediate results. XIX. An overview of the polarized thermal
  emission from Galactic dust},'' {\em A\&A, submitted,
  [arXiv:astro-ph/1405.0871]}  (2014).

\bibitem{planck2014-XX}
{\sorthelp{Planck Collaboration Int T}}{Planck Collaboration Int. XX},
  ``{\Planck\ intermediate results. XX. Comparison of polarized thermal
  emission from Galactic dust with simulations of MHD turbulence},'' {\em A\&A,
  submitted, [arXiv:astro-ph/1405.0872]}  (2014).

\bibitem{Hildebrand2009}
{Hildebrand}, R.~H., {Kirby}, L., {Dotson}, J.~L., {Houde}, M., and
  {Vaillancourt}, J.~E., ``{Dispersion of Magnetic Fields in Molecular Clouds.
  I},'' {\em ApJ}~{\bf 696},  567--573 (May 2009).

\bibitem{finkbeiner1999}
{Finkbeiner}, D.~P., {Davis}, M., and {Schlegel}, D.~J., ``{Extrapolation of
  Galactic Dust Emission at 100 Microns to Cosmic Microwave Background
  Radiation Frequencies Using FIRAS},'' {\em ApJ}~{\bf 524},  867--886 (Oct.
  1999).

\bibitem{Meny2007}
{Meny}, C., {Gromov}, V., {Boudet}, N., {Bernard}, J., {Paradis}, D., and
  {Nayral}, C., ``{Far-infrared to millimeter astrophysical dust emission. I. A
  model based on physical properties of amorphous solids},'' {\em A\&A}~{\bf
  468},  171--188 (June 2007).

\bibitem{planck2013-p01}
{\sorthelp{Planck Collaboration 2013A}}{Planck Collaboration I},
  ``{\textit{Planck} 2013 results: Overview of Planck Products and Scientific
  Results},'' {\em A\&A, in press}  (2014).

\bibitem{planck2013-XVII}
{\sorthelp{Planck Collaboration Int Q}}{Planck Collaboration Int. XVII},
  ``{Planck intermediate results. XVII. Emission of dust in the diffuse
  interstellar medium from the far-infrared to microwave frequencies},'' {\em
  A\&A, in press}  (2014).

\bibitem{planck2013-p06b}
{\sorthelp{Planck Collaboration 2013K}}{Planck Collaboration XI},
  ``{\textit{Planck} 2013 results: All-sky model of thermal dust emission},''
  {\em A\&A, in press}  (2014).

\bibitem{Molinari2010}
{Molinari}, S., {Swinyard}, B., {Bally}, J., {Barlow}, M., {Bernard}, J.-P.,
  {Martin}, P., {Moore}, T., {Noriega-Crespo}, A., {Plume}, R., {Testi}, L.,
  {Zavagno}, A., and {et al.}, ``{Hi-GAL: The Herschel Infrared Galactic Plane
  Survey},'' {\em PASP}~{\bf 122},  314--325 (Mar. 2010).

\bibitem{Engel2013}
{Engel}, C., {Ristorcelli}, I., {Bernard}, J.-P., {Longval}, Y., {Marty}, C.,
  {Mot}, B., {Otrio}, G., and {Roudil}, G., ``{Characterization and
  performances of the primary mirror of the PILOT balloon-borne experiment},''
  {\em Experimental Astronomy}~{\bf 36},  21--57 (Aug. 2013).

\end{thebibliography}
\bibliographystyle{spiebib}   

\end{document}